\documentclass[tighten,twocolumn,apj]{openjournal}

\usepackage{amsmath, bm, color, colortbl, multirow, xcolor, soul}
\usepackage[T1]{fontenc}

\definecolor{xlinkcolor}{cmyk}{1,1,0,0}

\usepackage[colorlinks=true, linkcolor=blue, citecolor=xlinkcolor]{hyperref}
\newcommand{\out}[1]{\textcolor{black}{#1}}

\definecolor{lightblue}{rgb}{.90,.95,1} \sethlcolor{lightblue}

\shorttitle{Microwave Spectra of Protoplanetary Disks}
\shortauthors{Painter et al.}
 
\begin{document}

\title{Detailed Microwave Continuum Spectra from Bright Protoplanetary Disks in Taurus}


\author{Caleb Painter\altaffilmark{1}, Sean M.~Andrews\altaffilmark{1}, Claire J.~Chandler\altaffilmark{2}, Takahiro Ueda\altaffilmark{1}, David J.~Wilner\altaffilmark{1}, \\ Feng Long\altaffilmark{3,11}, Enrique Mac{\'\i}as\altaffilmark{4}, Carlos Carrasco-Gonz{\'a}lez\altaffilmark{5}, Chia-Ying Chung\altaffilmark{6}, \\ Hauyu Baobab Liu\altaffilmark{6,7}, Tilman Birnstiel\altaffilmark{8,9}, and A.~Meredith Hughes\altaffilmark{10}}

\altaffiltext{1}{Center for Astrophysics \textbar\ Harvard \& Smithsonian, 60 Garden Street, Cambridge, MA 02138, USA}
\altaffiltext{2}{National Radio Astronomy Observatory, P.O. Box O, Socorro, NM 87801, USA}
\altaffiltext{3}{Lunar and Planetary Laboratory, University of Arizona, Tucson, AZ 85721, USA}
\altaffiltext{4}{European Southern Observatory, Karl-Schwarzschild-Str.~2, 85748, Garching bei M\"unchen, Germany}
\altaffiltext{5}{Instituto de Radioastronom{\'\i}a y Astrof{\'\i}sica (IRyA-UNAM), Morelia, Mexico}
\altaffiltext{6}{Department of Physics, National Sun Yat-Sen University, No.~70, Lien-Hai Road, Kaohsiung City 80424, Taiwan, R.O.C.}
\altaffiltext{7}{Center of Astronomy and Gravitation, National Taiwan Normal University, Taipei 116, Taiwan}
\altaffiltext{8}{University Observatory, Faculty of Physics, Ludwig-Maximilians-Universit\"at M\"unchen, Scheinerstr.~1, 81679 Munich, Germany}
\altaffiltext{9}{Exzellenzcluster ORIGINS, Boltzmannstr.~2, 85748 Garching, Germany}
\altaffiltext{10}{Department of Astronomy, Van Vleck Observatory, Wesleyan University, 96 Foss Hill Dr., Middletown, CT 06459, USA}
\altaffiltext{11}{NASA Hubble Fellowship Program Sagan Fellow.}

\begin{abstract}

We present new observations that densely sample the microwave (4--360 GHz) continuum spectra from eight young systems in the Taurus region.  Multi-component, empirical model prescriptions were used to disentangle the contributions from their dust disks and other emission mechanisms.  We found partially optically thick, free-free emission in all these systems, with positive spectral indices (median $\alpha_{\rm c} \approx 1$ at 10 GHz) and contributing 5--50\%\ of the 43 GHz fluxes.  There is no evidence for synchrotron or spinning dust grain emission contributions for these targets.  The inferred dust disk spectra all show substantial curvature: their spectral indices decrease with frequency, from $\alpha_{\rm d} \approx 2.8$--4.0 around 43 GHz to 1.7--2.1 around 340 GHz.  This curvature suggests that a substantial fraction of the (sub)millimeter ($\gtrsim$\,\,200 GHz) dust emission may be optically thick, and therefore the traditional metrics for estimating dust masses are flawed.  Assuming the emission at lower frequencies (43 GHz) is optically thin, the local spectral indices and fluxes were used to constrain the disk-averaged dust properties and estimate corresponding dust masses.  These masses are roughly an order of magnitude higher ($\approx 1000 \, M_\oplus$) than those found from the traditional approach based on (sub)millimeter fluxes.  These findings emphasize the value of broad spectral coverage -- particularly extending to lower frequencies ($\sim$cm-band) -- for accurately interpreting dust disk emission; such observations may help reshape our perspective on the available mass budgets for planet formation.

\end{abstract}

\section{Introduction} \label{sec:intro}

The planet formation process depends critically on the evolution of the solid mass available in the progenitor circumstellar disk \citep[e.g.,][]{pollack96,drazkowska23}.  Information about disk solids is accessed through the microwave ($\sim$20--500 GHz) thermal continuum emitted by dust \citep{andrews20}: if optical depths are low, the continuum flux density ($S_\nu$) scales with mass \citep{beckwith90}, and the spectral index ($\alpha$, where $S_\nu \propto \nu^\alpha$) traces the shape of the opacity spectrum $\kappa_\nu$ (where $\kappa_\nu \propto \nu^\beta$, and $\alpha \approx 2 + \beta$ in the Rayleigh-Jeans limit; \citealt{beckwith91,ricci10a}) and thereby the dust size distribution \citep{miyake93,dalessio06,draine06}.  ALMA surveys have constrained these fundamental metrics for planet formation by tracking how $S_\nu$ and $\alpha$ evolve and depend on other properties \citep[see][]{manara23}.

For standard assumptions, the observed $S_\nu$ distributions for nearby $\sim$1--3 Myr-old disks imply that there is not enough progenitor mass available (by $\gtrsim$\,3$\times$) to explain the observed exoplanet population \citep{manara18}, driving speculation that planet formation finished early \citep{greaves10,najita14,tychoniec20} or that mass is externally replenished \citep{throop08,winter24}.  A more mundane possibility is that a significant fraction of $S_\nu$ -- usually measured at 230/340 GHz (ALMA Bands 6/7) -- is optically thick \citep{beckwith90,aw05}.  In optically thick regions, brightness temperatures saturate at the local dust temperatures, or lower if the albedos are high \citep{miyake1993effects,dsharp5,zhu19}.  In this case, $S_\nu$ is insensitive to mass (it sets a lower limit) and $\alpha$ is insensitive to particle sizes ($\alpha \approx 2$; \citealt{liu19}).

\begin{deluxetable*}{l c c c c | c c c}[ht!]
\tabletypesize{\footnotesize}
\tablecaption{Observing Log \label{table:obs_log}}
\tablehead{
\colhead{UTC Date and Time} & \colhead{Facility} & \colhead{Spectral Settings} & \colhead{Config.} & \colhead{Target(s)} & \colhead{Gain} & \colhead{Bandpass} & \colhead{Flux}
}
\startdata
2022/07/28/12:18--14:50 & VLA & C,\,X,\,Ku,\,K,\,Ka,\,Q & D & GM Aur   & J0439+3045  & 3C\,138     & 3C\,138   \\
2022/08/06/18:49--19:42 & VLA & C,\,X,\,Ku,\,K,\,Ka,\,Q & D & (3C\,138) & (3C\,138)    & (3C\,286)   & (3C\,286) \\
2022/08/10/09:55--12:26 & VLA & C,\,X,\,Ku,\,K,\,Ka,\,Q & D & CI Tau   & J0426+2327  & 3C\,138     & 3C\,138   \\
2022/08/12/09:47--12:18 & VLA & C,\,X,\,Ku,\,K,\,Ka,\,Q & D & LkCa 15  & J0426+2327  & 3C\,138     & 3C\,138   \\
2022/08/13/09:55--12:27 & VLA & C,\,X,\,Ku,\,K,\,Ka,\,Q & D & MWC\,480  & J0439+3045  & 3C\,138     & 3C\,138   \\
2022/08/14/09:34--12:06 & VLA & C,\,X,\,Ku,\,K,\,Ka,\,Q & D & RY Tau   & J0439+3045  & 3C\,138     & 3C\,138   \\
2022/08/15/09:47--12:18 & VLA & C,\,X,\,Ku,\,K,\,Ka,\,Q & D & DL Tau   & J0426+2327  & 3C\,138     & 3C\,138   \\
2022/08/15/12:18--14:49 & VLA & C,\,X,\,Ku,\,K,\,Ka,\,Q & D & DR Tau   & J0440+1437  & 3C\,138     & 3C\,138   \\
2022/08/19/10:26--12:57 & VLA & C,\,X,\,Ku,\,K,\,Ka,\,Q & D & FT Tau   & J0426+2327  & 3C\,138     & 3C\,138   \\
2022/08/28/17:17--18:10 & VLA & C,\,X,\,Ku,\,K,\,Ka,\,Q & D & (3C\,138) & (3C\,138)    & (3C\,286)   & (3C\,286) \\
\hline
2023/04/10/12:38--16:44 & NOEMA & Setting 2 (Band 1) & 12C & all & J0438+300 & 3C\,84    & LkH$\alpha$ 101 \\   
2023/04/19/15:22--20:27 & NOEMA & Setting 1 (Band 1) & 12C & all & J0438+300 & 3C\,84    & LkH$\alpha$ 101 \\   
2023/04/22/11:36--14:34 & NOEMA & Setting 2 (Band 1) & 12C & all & J0438+300 & 3C\,454.3 & MWC 349         \\   
2023/04/22/15:02--18:10 & NOEMA & Setting 2 (Band 1) & 12C & all & J0438+300 & 3C\,454.3 & MWC 349         \\   
2023/10/08/00:29--06:48 & NOEMA & Setting 3 (Band 2) & 10C & all & J0438+300 & 3C\,84    & MWC 349         \\   
2023/10/09/01:04--05:57 & NOEMA & Setting 4 (Band 2) & 10C & all & J0438+300 & 3C\,84, 3C\,454.3 & MWC 349 \\ 
\hline
2023/01/22/07:03--12:10 & SMA   & Setting 1 (Rx230/240) & SUB & all & 3C\,111, J0510+180   & 3C\,273          & Uranus \\
2023/02/08/02:54--11:55 & SMA   & Setting 2 (Rx240/345) & SUB & all & 3C\,111, J0510+180   & J1159+292       & Uranus \\
2023/02/09/04:33--11:25 & SMA   & Setting 3 (Rx345/400) & SUB & all & 3C\,111, J0510+180   & J1159+292       & Uranus
\enddata
\end{deluxetable*}

There is a range of compelling evidence that high optical depths at higher frequencies ($\nu \gtrsim 100$ GHz) can bias disk mass estimates \citep[e.g.,][]{xin23}.  At high resolution, the (sub)mm continuum often shows high brightness temperatures \citep{dsharp2} but shallow spectra ($\alpha \approx 2$; \citealt{tsukagoshi16,huang18,long20,guidi22}), particularly within a few tens of au of the host stars and in small-scale substructures.  At coarser resolution, larger samples reveal a size--luminosity scaling where $S_\nu$ correlates with surface area \citep{tripathi17}, as expected for high optical depths.  And recently, there is emerging evidence that continuum spectra are steeper at lower, optically thinner, frequencies \citep{chung25,garufi25}, bolstering the idea that (sub)mm-band emission (with mean $\alpha_{\rm mm} \approx 2.2$; \citealt{tazzari21,chung24}) is ``polluted'' by high optical depths.

If such pollution is indeed universal, then current estimates of disk solid masses based on (sub)mm continuum emission are flawed, reflecting instead some more ambiguous (and biased) combination of optically thick fractions, sizes, and heating.  Accurate assessments of these masses, and thereby the potential for planet formation, require access to optically thin emission.  Given the potential for high optical depths at $\nu \gtrsim 100$ GHz, the only viable approach is to extend the continuum measurements to lower frequencies, where optical depths are diminished \citep{wilner2004imaging,rodmann06}.  However, this is not without some added challenges, since we also find significant contributions from non-dust emission at these lower frequencies.  Usually, that ``contamination'' spectrum is sparsely sampled, introducing considerable ambiguity into the interpretations of the composite spectrum.  In this article, we present sensitive new observations that densely sample the complete microwave spectra for a set of nearby, bright disks, with the aim of more precisely constraining the shapes of their optically thinner, $\sim$cm-wavelength dust emission.  Section 2 describes the observations and their calibration.  Section 3 explains the spectrum measurements and modeling formalism, and Section 4 details the results and corresponding mass estimates.  Section 5 discusses the implications of this analysis in the broader landscape of disk mass estimates, and speculates on potentially fruitful new avenues for this field with future observations.

\section{Observations and Data Reduction} \label{sec:obs}

The goal in this study was to measure the continuum emission from young stars and their disks across a wide swath of frequencies ($\sim$4--400 GHz) with sufficient sensitivity to densely probe the spectrum with a fine frequency sampling (at 1--2 GHz). Technical improvements that provide increased instantaneous bandwidth at the VLA, NOEMA, and SMA now make it possible to assemble densely sampled spectra of this nature.  To that end, we selected eight target systems in the nearby Taurus star-forming region known to be especially bright in the 230 or 340 GHz bands from previous measurements: CI Tau, DL Tau, DR Tau, FT Tau, GM Aur, LkCa 15, MWC 480, and RY Tau \citep{beckwith90,mannings97,aw05}.  These same targets have the benefit of existing high angular resolution measurements for at least one microwave frequency \citep{clarke18,long18,huang20}, which can be leveraged along with the spectra analyzed here to provide robust, combined constraints on the physical characteristics of the disk solids in future work.

\begin{figure*}[ht!]
    \centering
    \includegraphics[width=\linewidth]{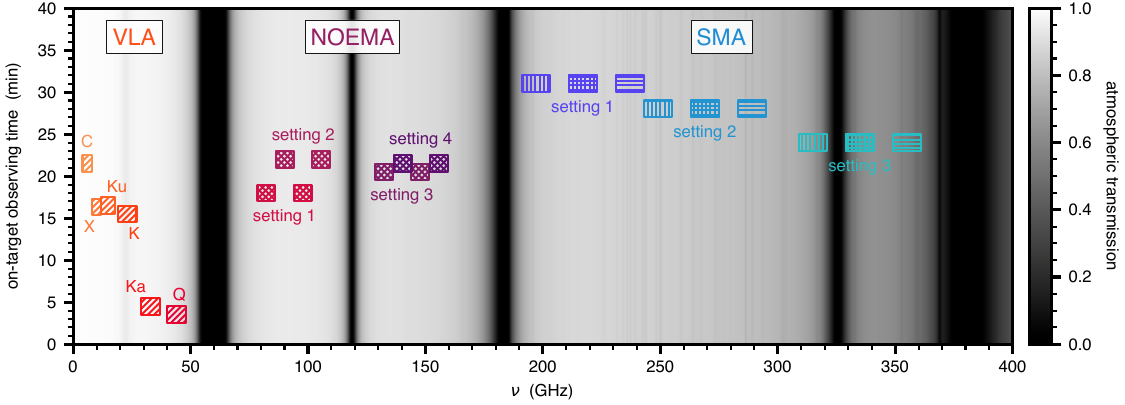}
    \caption{A visualization of the spectral coverage and (cumulative) on-target integration times.  See Table \ref{table:spectral_setups} for more information on the spectral settings.  The background grayscale shows a representative composite spectrum of the atmospheric transmission, stitched together from {\tt am} models \citep{paine_am} for mean conditions at the respective site altitudes.}
    \label{fig:spectral_settings}
\end{figure*}

Acquiring the desired breadth in frequency coverage for this project involved observations in multiple settings with each of three different interferometers.  The remainder of this section describes the observational design and calibration of the data for each of these facilities.  A combined observing log is provided in Table \ref{table:obs_log}, and a summary of the spectral configurations is available in Table \ref{table:spectral_setups}.  Figure \ref{fig:spectral_settings} offers some visual context for the frequency coverage obtained from these programs.

\begin{deluxetable*}{l c c c r | c}[ht!]
\tabletypesize{\footnotesize}
\tablecaption{Spectral Setups \label{table:spectral_setups}}
\tablehead{
\colhead{Facility} & \colhead{Setting} & \colhead{Rx} & \colhead{Frequency Coverage (GHz)} & \colhead{SPWs (\# $\times$ $\Delta \nu$ @ chans)} & \colhead{Sub-band frequencies (GHz)}
}
\startdata
\multirow{6}{*}{VLA}   & C  & C             & 4--6, 6--8                              & $32 \times 128$ MHz @ 2 MHz & 4.6, 5.4, 6.6, 7.4 \\
                       & X  & X             & 8--10, 10--12                           & $32 \times 128$ MHz @ 2 MHz & 8.5, 9.5, 10.5, 11.4 \\
                       & Ku & Ku            & 12--14, 14--16, 16--18                  & $48 \times 128$ MHz @ 2 MHz & 13.0, 14.3, 15.8, 17.3 \\
                       & K  & K             & 18--20, 20--22, 22--24, 24--26          & $64 \times 128$ MHz @ 2 MHz & 19.0, 20.7, 23.1, 25.0 \\
                       & Ka & Ka            & 29--31, 31--33, 33--35, 35--37          & $64 \times 128$ MHz @ 2 MHz & 30, 32, 34, 36 \\
                       & Q  & Q             & 40--42, 42--44, 44--46, 46--48          & $64 \times 128$ MHz @ 2 MHz & 41, 43, 45, 47 \\
\hline
\multirow{4}{*}{NOEMA} & 1  & Band 1        & 78.3--86.2, 93.8--101.7                 & $4 \times 3.85$ GHz @ 2 MHz & 80, 84, 96, 100 \\ 
                       & 2  & Band 1        & 86.3--94.2, 101.8--109.7                & $4 \times 3.85$ GHz @ 2 MHz & 88, 92, 104, 108 \\
                       & 3  & Band 2        & 128.3--136.4, 143.8--151.7              & $4 \times 3.85$ GHz @ 2 MHz & 130, 134, 146, 150 \\
                       & 4  & Band 2        & 136.3--144.2, 151.8--159.7              & $4 \times 3.85$ GHz @ 2 MHz & 138, 142, 154, 158 \\
\hline
\multirow{3}{*}{SMA}   & 1  & Rx230 / Rx240 & 191--203, 211--223 / 211--223, 231--243 & $24 \times 2$ GHz @ 140 kHz & 194, 200, 214, 220, 234, 240 \\
                       & 2  & Rx240 / Rx345 & 243--255, 263--275 / 263--275, 283--295 & $24 \times 2$ GHz @ 140 kHz & 246, 252, 266, 272, 286, 292 \\
                       & 3  & Rx345 / Rx400 & 309--321, 329--341 / 329--341, 349--361 & $24 \times 2$ GHz @ 140 kHz & 314, 336, 356 
\enddata
\tablecomments{The VLA and NOEMA measurements are dual-polarization, meaning the total correlated bandwidth is $2\times$ what is specified above (with the same frequency coverage).  The two polarization states were averaged in our analysis.  The SMA measurements split different linear polarization states to two receivers with different spectral settings.  The mean frequencies of the `sub-bands' (rightmost column) are approximate; small shifts owing to the specific weighted averages were accounted for in the analysis.}
\end{deluxetable*}

\subsection{VLA}

We conducted ten observing sessions with the 27 antennas (25 m diameter) of the Karl G.~Jansky Very Large Array (VLA), in its D configuration with baseline lengths of 35--1030 m (project {\tt 22A--179}).  In eight of those 2.5 hour sessions, we cycled through observations of a single disk target (and relevant calibrators) in six VLA receiver bands: C, X, Ku, K, Ka, and Q.  The on-target integration times ranged from 3 to 23 minutes (see Fig.~\ref{fig:spectral_settings}).  Although this strategy is relatively inefficient (due to the overheads of switching receiver bands) compared to cycling through all the sources in a single band per session, it has the advantage of eliminating confusion by any potential source variability on $\gtrsim$\,1 hour timescales.  The remaining two sessions observed both the nearby flux calibrator from the other sessions (3C 138) and the primary flux calibrator 3C 286 in the same sequence of receiver bands, to enable a `bootstrap' of the flux scale from the flaring 3C 138 to a stable calibrator (as described below).  Given their locations, the disk targets and 3C 286 cannot be observed together in the same scheduling block.  In the first of those bootstrap sessions (2022/08/06), the conditions were not suitable for measurements at the highest frequency bands (Ka and Q).  Table \ref{table:obs_log} records the observing dates, times, and other relevant details.

All of these observations used a single correlator setting for each dual-polarization receiver, with the 3-bit samplers optimized for continuum sensitivity.  Each setting deployed adjacent 2 GHz-wide basebands, with each baseband divided into 16 spectral windows (SPWs) that covered 128 MHz with $64 \times 2$ MHz channels (in each polarization).  The Q, Ka, and K observations used four basebands (8 GHz bandwidth); Ku used three (6 GHz); X and C used two (4 GHz).  The spectral settings are summarized in Table \ref{table:spectral_setups} (and Fig.~\ref{fig:spectral_settings}).

\begin{figure}[th!]
    \centering
    \includegraphics[width=\linewidth]{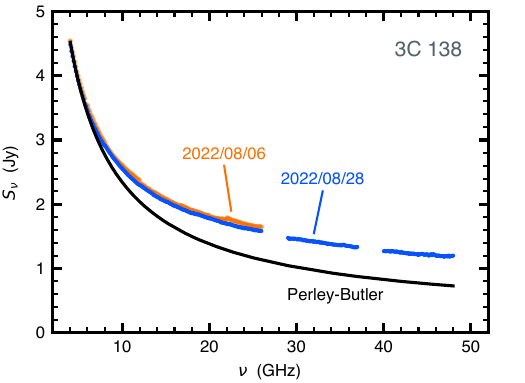}
    \caption{The measured VLA spectrum for 3C 138, bootstrapped from joint observations with 3C 286 on two occasions, compared to the \citet{perley17} model.  The flaring 3C 138 spectrum does not substantially change during the $\sim$3 week observing timeframe.  We adopted the 2022/08/28 measurement to correct the flux calibration scale from the standard VLA pipeline outputs.}
    \label{fig:rescale_VLA}
\end{figure}

The first steps in reducing these data were the standard flagging and other corrections, bandpass, gain, and flux calibration using the VLA pipeline ({\tt v2023.1.0.124}) in {\sf CASA} \citep{CASA22}.  Next, we corrected the flux calibration scale to account for the flaring 3C 138 during this period.  We used the bootstrap observing sessions to determine the deviation of the 3C 138 spectrum from the pipeline-default \citet{perley17} model, by referencing the measurements to the observed 3C 286 emission (and model).  Those measurements and monitoring at the VLA (project {\tt TCAL0009}) demonstrated that flare variations over the observing period for this program were small.  The 3C 138 rescaling factors we used for each SPW were derived from the spectra presented in Figure \ref{fig:rescale_VLA}: we adopted the measurements from the observations on 2022/08/28.  These were applied as gain corrections to each dataset.  We conducted some occasional additional manual flagging to excise contamination from radio frequency interference (in X and C bands) and then averaged all the channels in each SPW.

Next, we used {\sf CASA} to image the data in each receiver band with natural weighting, using the multiterm multifrequency synthesis algorithm ({\tt MTMFS}, {\tt nterms=2}; \citealt{rau11}) in the {\tt tclean} task.  We then used those {\tt tclean} models as the starting points for self-calibrating the phases on scan-length intervals (combining SPWs), if there was sufficient signal-to-noise (SNR $> 3$).  When possible, this offered a modest improvement in the peak SNR in re-imaged maps.  This process for the lower frequency bands (particularly C) was more complicated and tailored to individual datasets, due to the varied presence of bright background sources in the fields.  We followed a standard `peeling' prescription for bright sources outside the primary beam, using self-calibration phase and amplitude solutions (sometimes on a per SPW basis) to iteratively define and then subtract models of their emission.  Then, for cases with background sources closer to the phase center, we took advantage of their presence to derive improved phase and amplitude self-calibration solutions for the disk target region.

The final step was to spectrally split out the fully calibrated visibilities in a set of `sub-bands' for further analysis (Sect.~\ref{sec:analysis}).  For the K, Ka, and Q bands, these sub-bands were the same as the 2 GHz-wide basebands (16 SPWs each).  To minimize the impact of spectral gradients across the sub-bands at lower frequencies, we used four (contiguous) 1.5 GHz-wide (Ku; 12 SPWs each) or 1 GHz-wide (X, C; 8 SPWs each) sub-bands.  When necessary in the lower frequency bands (C, X, Ku), we iteratively subtracted point source models for each sub-band to remove the emission from all background sources before further analysis.  The approximate weighted mean frequencies in each sub-band are listed in Table \ref{table:spectral_setups}.

\subsection{NOEMA}

We also observed in six sessions (Table \ref{table:obs_log}) with ten to twelve of the 15 m antennas that comprise the IRAM Northern Extended Millimeter Array (NOEMA), in the 10C or 12C configurations that cover 24--368 m baselines (project {\tt W22AX}).  The eight targets were observed together in each session, with 30--40 min cycles bracketed by the gain calibrator(s).  Bandpass and flux calibrators were observed in each session.  The cumulative on-source integration times for each target were 18--23 min.

Four spectral settings were configured (Table \ref{table:spectral_setups}), two each for the dual-polarization, dual-sideband Band 1 and 2 receivers.  In each setting, the wideband PolyFiX correlator was configured to cover 15.5 GHz in two polarizations, divided into two sidebands (equivalent basebands) separated by 8 GHz, with each split into two 3.875 GHz-wide SPWs sampled with 2 MHz channels.  As illustrated in Figure \ref{fig:spectral_settings}, these settings were designed to `nest', so the lower sideband of the higher frequency setting for each receiver covers the gap between the sidebands of the lower frequency setting.  The results are 31 GHz of contiguous coverage in each band.  This nesting improves the relative flux calibration of neighboring settings (see below).  

The visibilities were calibrated with the {\sf GILDAS CLIC} pipeline, including routine tasks like flagging, bandpass, gain, and flux calibrations, as well as delay and polarization corrections.  Then, the calibrated visibilities were spectrally averaged in each SPW before exporting for further reduction in {\sf CASA}.  We first constructed naturally-weighted images of each SPW (combining polarizations) using the {\tt MTMFS} algorithm in {\tt tclean}.  These were used as starting points for an iterative self-calibration, with phase-only and then amplitude+phase solutions on decreasing intervals ($\sim$180, 90, 30 s) when SNR $> 5$.  This self-calibration significantly improved the data quality, with peak SNRs in re-imaged maps increasing 2--10$\times$.     

We then split the calibrated visibilities into sub-bands equivalent to the SPWs, averaging both polarizations, for further analysis (Table \ref{table:spectral_setups}).  Following Section \ref{sec:spec_measure}, we derived a first pass at flux density measurements for each sub-band and inspected the spectra for all targets to search of mismatches between the nested settings.  We found agreement between Settings 3 and 4, but a persistent offset 7\%\ high for Setting 1 compared to 2.  A gain correction was applied to the Setting 1 data to compensate, but we note that such calibration systematics will be considered further for the analysis in Section \ref{sec:analysis}.

\subsection{SMA}

Finally, we observed in three sessions with the Submillimeter Array (SMA; \citealt{ho04}), using six or seven of the 6 m antennas in the sub-compact configuration with 9--70 m baselines (project {\tt 2022B--S015}).  Each session cycled through all eight targets every 32 minutes, interleaved with the gain calibrators.   Since Uranus was relatively nearby ($\sim$25\degr\ to the SE), it was observed throughout each track as a primary flux calibrator.  The bandpass calibrator was measured at the end of each track. 

These observations directed orthogonal polarizations to two distinct dual-sideband receivers, with the same correlator settings but shifted frequency coverages (Table \ref{table:spectral_setups}).  We processed 12 GHz of bandwidth for each receiver and sideband (with an 8 GHz separation between sidebands), split into six 2 GHz SPWs sampled with 140 kHz channels.  The setups were designed so that the upper sideband of one receiver overlaps with the lower sideband of the other receiver (Fig.~\ref{fig:spectral_settings}), to minimize systematics in their relative flux calibrations (see below).  The result is 96 GHz of coverage spanning the more transparent parts of the atmosphere from $\sim$190--360 GHz. 

The visibility data were calibrated with the {\sf MIR} facility software, including standard tasks of despiking and flagging, system temperature correction, bandpass and gain calibration, and spectral averaging.  The Uranus observations were used to derive the flux scale independently in each 2 GHz SPW, since {\sf MIR} does not properly scale the calibrator models with frequency.  The calibrated visibilities were ingested into {\sf CASA} format, and then the data from each receiver in a given setting was imaged with natural weighting using the {\tt MTMFS} algorithm in {\tt tclean} to establish an initial image model.  We performed a phase-only self-calibration for the combined SPWs in each receiver/setting combination on decreasing solution intervals (240, 120, 60 s) when SNR $> 5$, finding peak SNR improved by $\sim$1.3--2.5$\times$ in re-imaged maps (we found no benefits from amplitude+phase self-calibration).   

For Settings 1 and 2, we next averaged the data into 6 GHz-wide sub-bands (3 SPWs each) and measured preliminary flux densities, following Section \ref{sec:spec_measure}.  We compared the spectra for the two sub-bands in each setting that fully overlap (one from each receiver/polarization), and found minor, but systematic, mismatches.  To `align' the spectra, we applied a gain correction to decrease the Rx230 amplitude scales in Setting 1 by 3\%, and increase the Rx240 scales in Setting 2 by 5\%.  We then combined the data from the overlapping sub-bands, resulting in six effective sub-bands per setting (Table \ref{table:spectral_setups}).  For Setting 3, a closer examination of the data showed that the SPWs at the inner edges of both sidebands for Rx345 and the low-frequency edge of the lower sideband for Rx400 were substantially noisier than their counterparts as a result of their proximity to the 325 GHz atmospheric H$_2$O feature (Fig.~\ref{fig:spectral_settings}).  They were excised, the self-calibration was repeated, and then we compared the overlapping sub-bands as above.  We found no significant differences in the overlap.  To match the typical SNR in the lower frequency settings, we chose to combine the (noisier) Setting 3 data into three broader (10--12 GHz) sub-bands (Table \ref{table:spectral_setups}).

\subsection{Ancillary Literature Data} \label{sec:lit}

To add some context for these new data, we aggregated microwave continuum photometry measurements for the target disks from the literature.  These include measurements at low frequencies to compare with the VLA data \citep{rodmann06,ricci10a,isella14,dzib15,macias16,tazzari16,zapata17,espaillat19,greaves22,garufi25,chung25}; at intermediate frequencies that overlap with the NOEMA data \citep{dutrey96,ohashi96,looney00,duvert00,kitamura02,pietu06,guilloteau11,kwon15,harrison19}; at the more commonly observed higher frequencies probed with the SMA data \citep{weintraub89,beckwith90,adams90,beckwith91,altenhoff94,mannings94,mannings97,osterloh95,aw05,aw07a,hamidouche06,hughes08,hughes09,isella09,isella12,oberg10,andrews11,sandell11,tripathi17,huang17,huang20,long18,long19,sturm22,harrison24,chung24}; and extending to some even higher ($\sim$THz) frequencies \citep{beichman88,ishihari10,rebull10,howard13,ribas17}.  These compilations are made available as machine-readable files at \url{https://zenodo.org/records/16322048}.

\section{Analysis} \label{sec:analysis}

\subsection{Spectrum Measurements} \label{sec:spec_measure}

We measured the spectrum for each target by forward-modeling the (combined) visibilities for each sub-band, similar to the approach of \citet{chung24}.  For a set of $N$ (complex) visibilities $\mathcal{V}$ and weights $w$ that define the data, and a corresponding set of model visibilities $\mathsf{V}$ that depend on parameters $\theta$, we defined a standard Gaussian likelihood function
\begin{equation}
    \ln {\sf p}(\mathcal{V} | \, \theta) \propto - \frac{1}{2} \sum_j^N \left( \frac{| \mathcal{V}_j - \mathsf{V}_j(\theta)|^2}{s_j^2} + \ln{2\pi s_j^2} \right),
    \label{eq:photometry_likelihood}
\end{equation}
where $s_j^2 = w_j^{-1} + f_S^2 |\mathsf{V}_j|^2$ is a variance term that accounts for both the visibility weights and an additional amount of scatter defined as a fraction $f_S$ of the model amplitude (where $f_S$ is treated as a free parameter).  The latter contribution is designed to mitigate the potential bias in inferences (particularly in $S_\nu$) caused by any additional scatter in the data \citep[e.g.,][]{tremaine02}.  In this analysis, we considered data with projected baselines $\le65$k$\lambda$ only, to avoid having to adopt more complex models (this only practically affects some NOEMA data, and is negligible for the flux measurements).  For each dataset, we re-scaled the visibility weights based on comparisons of their means to the measured dispersions in the imaginary components.  We retrospectively confirmed these scalings based on the dispersions of the residuals from the best-fit models.      

We considered two emission models, adopting a point source for $\nu < 18$ GHz and an elliptical Gaussian otherwise.  The point source model has parameters $\theta = [S_\nu, \Delta x, \Delta y, f_S]$, where $S_\nu$ is flux and $\Delta x$, $\Delta y$ are right ascension and declination offsets, and is defined as
\begin{equation}
    \mathsf{V}^{\tt point} = S_\nu e^{2 \pi i (u \Delta x + v \Delta y) b},
\end{equation}
where $u, v$ are the Fourier spatial frequency coordinates (in wavelengths) and $b = (\pi / 180) / 3600$ converts the offsets from arcseconds to radians.  The Gaussian model has parameters $\theta = [S_\nu, \Delta x, \Delta y, \Delta R, f_S]$, where $\Delta R$ is the (deprojected) standard deviation, and is defined as
\begin{equation}
    \mathsf{V}^{\tt gauss} = \mathsf{V}^{\tt point} e^{-2\pi^2(u_r^2 + v_r^2 \mu^2) \Delta R^2 b^2},
\end{equation}
where $\mu$ is the cosine of the disk inclination angle and the rotated Fourier coordinates are
\begin{eqnarray}
    u_r &=& \phantom{-}u \sin{\vartheta} + v\cos{\vartheta} \\
    v_r &=& -u \sin{\vartheta} + v\sin{\vartheta},
\end{eqnarray}
with $\vartheta$ the position angle of the disk major axis.  Given the limited resolution of these data, we fixed $\mu$ and $\vartheta$ to the values measured at higher resolution in the literature (\citealt{jennings22} for CI Tau and DL Tau; \citealt{long18} for FT Tau and RY Tau; \citealt{huang20} for GM Aur; \citealt{huang23} for DR Tau; \citealt{exoALMA_I} for LkCa 15; and \citealt{andrews24} for MWC 480).

We assigned uniform priors for $S_\nu$ (permitting negative values), $\Delta R$, and $\ln{f_S}$, and Gaussian priors for the offsets that were centered around the peak emission locations in synthesized images and had standard deviations of 1\arcsec\ in each direction.  The posterior distributions were sampled using {\tt emcee} \citep{foreman-mackey13}, which builds on the \citet{goodman10} MCMC ensemble sampler, deploying 64 walkers initialized with random draws from the priors and running for 10,000 steps.  We discarded 10$\times$ the autocorrelation times ($\lesssim$\,100 steps) as burn-in.  For this study, we are only interested in the marginalized posteriors for $S_\nu$.

\begin{deluxetable*}{c | c c c c c c c c}[t!]
\tabletypesize{\scriptsize}
\tablecaption{Measured Microwave Spectra \label{table:spectra}}
\tablehead{
\colhead{} & \multicolumn{8}{c}{Measured Fluxes: $S_\nu \pm \sigma_S$ (mJy)} \\[1mm]
\colhead{$\nu$ (GHz)} & \colhead{CI Tau} & \colhead{DL Tau} & \colhead{DR Tau} & \colhead{FT Tau} & \colhead{GM Aur} & \colhead{LkCa 15} & \colhead{MWC 480} & \colhead{RY Tau}
}
\startdata
4.6 &  \hl{$0.005\pm0.015$} & $0.066\pm0.012$          & \hl{$0.004\pm0.013$} & \hl{$0.014\pm0.011$} & $0.052\pm0.016$          & \hl{$0.006\pm0.013$} & $0.105\pm0.017$ & $0.249\pm0.016$ \\
5.4 &  \hl{$0.021\pm0.014$} & \hl{$0.028\pm0.013$} & \hl{$0.006\pm0.015$} & $0.037\pm0.011$          & \hl{$0.031\pm0.014$} & \hl{$0.010\pm0.011$} & $0.113\pm0.016$ & $0.259\pm0.016$ \\
6.6 &  $0.054\pm0.013$          & $0.055\pm0.012$          & $0.042\pm0.014$          & $0.038\pm0.010$          & $0.061\pm0.018$          & \hl{$0.018\pm0.012$} & $0.150\pm0.015$ & $0.318\pm0.015$ \\
7.4 &  $0.044\pm0.012$          & $0.038\pm0.011$          & \hl{$0.025\pm0.011$} & $0.055\pm0.009$          & $0.074\pm0.014$          & \hl{$0.019\pm0.010$} & $0.169\pm0.013$ & $0.319\pm0.013$ \\
8.5 &  $0.074\pm0.018$          & $0.083\pm0.019$          & $0.056\pm0.018$          & $0.075\pm0.016$          & $0.083\pm0.017$          & $0.038\pm0.011$          & $0.197\pm0.016$ & $0.367\pm0.018$ \\
9.5 &  $0.093\pm0.019$          & $0.114\pm0.018$          & $0.060\pm0.019$          & $0.087\pm0.016$          & $0.092\pm0.019$          & $0.037\pm0.010$          & $0.233\pm0.014$ & $0.388\pm0.016$ \\
10.5 & $0.086\pm0.018$          & $0.113\pm0.020$          & $0.085\pm0.022$          & $0.104\pm0.016$          & $0.125\pm0.019$          & $0.046\pm0.011$          & $0.257\pm0.015$ & $0.419\pm0.017$ \\
11.4 & $0.105\pm0.022$          & $0.103\pm0.022$          & $0.069\pm0.023$          & $0.096\pm0.019$          & $0.156\pm0.016$          & $0.050\pm0.013$          & $0.292\pm0.017$ & $0.476\pm0.020$ \\
13.0 & $0.133\pm0.026$          & $0.147\pm0.023$          & $0.137\pm0.042$          & $0.119\pm0.018$          & $0.125\pm0.019$          & $0.084\pm0.018$          & $0.341\pm0.018$ & $0.575\pm0.022$ \\
14.3 & $0.139\pm0.020$          & $0.164\pm0.018$          & $0.159\pm0.021$          & $0.125\pm0.016$          & $0.156\pm0.016$          & $0.075\pm0.014$          & $0.367\pm0.014$ & $0.589\pm0.016$ \\
15.8 & $0.151\pm0.023$          & $0.171\pm0.019$          & $0.153\pm0.025$          & $0.160\pm0.016$          & $0.167\pm0.018$          & $0.103\pm0.016$          & $0.438\pm0.015$ & $0.678\pm0.017$ \\
17.3 & $0.195\pm0.030$          & $0.186\pm0.023$          & $0.200\pm0.031$          & $0.184\pm0.019$          & $0.165\pm0.021$          & $0.105\pm0.019$          & $0.479\pm0.018$ & $0.731\pm0.021$ \\
19.0 & $0.181\pm0.031$          & $0.230\pm0.034$          & $0.251\pm0.042$          & $0.209\pm0.033$          & $0.198\pm0.034$          & $0.157\pm0.029$          & $0.538\pm0.034$ & $0.837\pm0.033$ \\
20.7 & $0.228\pm0.050$          & $0.314\pm0.052$          & $0.275\pm0.052$          & $0.304\pm0.047$          & $0.207\pm0.052$          & $0.203\pm0.043$          & $0.608\pm0.052$ & $0.924\pm0.046$ \\
23.1 & $0.335\pm0.074$          & $0.332\pm0.068$          & $0.217\pm0.059$          & $0.312\pm0.067$          & $0.348\pm0.066$          & \hl{$0.188\pm0.063$} & $0.766\pm0.072$ & $1.119\pm0.068$ \\
25.0 & $0.255\pm0.047$          & $0.450\pm0.046$          & $0.361\pm0.050$          & $0.394\pm0.046$          & $0.370\pm0.043$          & $0.309\pm0.043$          & $0.828\pm0.051$ & $1.232\pm0.048$ \\
30   & $0.426\pm0.114$          & $0.716\pm0.081$          & $0.608\pm0.083$          & $0.574\pm0.071$          & $0.534\pm0.093$          & \hl{$0.202\pm0.082$} & $1.256\pm0.087$ & $1.707\pm0.077$ \\
32   & $0.397\pm0.121$          & $0.819\pm0.083$          & $0.682\pm0.088$          & $0.644\pm0.070$          & $0.661\pm0.089$          & $0.474\pm0.089$          & $1.424\pm0.094$ & $1.836\pm0.083$ \\
34   & $0.720\pm0.140$          & $0.954\pm0.091$          & $0.735\pm0.099$          & $0.881\pm0.074$          & $0.708\pm0.105$          & $0.356\pm0.100$          & $1.710\pm0.099$ & $2.020\pm0.087$ \\
36   & $0.552\pm0.144$          & $1.121\pm0.103$          & $0.857\pm0.115$          & $0.928\pm0.084$          & $0.821\pm0.112$          & $0.445\pm0.114$          & $1.875\pm0.113$ & $2.368\pm0.100$ \\
41   & $0.63\pm0.29$            & $1.55\pm0.21$            & $1.15\pm0.23$            & $1.19\pm0.21$            & $1.15\pm0.28$            & \hl{$0.56\pm0.23$}   & $2.77\pm0.30$   & $3.05\pm0.24$   \\
43   & $1.15\pm0.34$            & $1.73\pm0.25$            & $1.76\pm0.26$            & $1.74\pm0.22$            & $1.34\pm0.28$            & \hl{$0.73\pm0.27$}   & $2.97\pm0.36$   & $3.44\pm0.27$   \\
45   & \hl{$0.83\pm0.38$}   & $2.13\pm0.35$            & $1.75\pm0.36$            & $2.01\pm0.30$            & $1.45\pm0.45$            & \hl{$1.00\pm0.35$}   & $3.20\pm0.46$   & $3.71\pm0.38$   \\
47   & \hl{$0.76\pm0.36$}   & $2.14\pm0.39$            & $1.98\pm0.46$            & $2.14\pm0.36$            & $2.15\pm0.47$            & $1.55\pm0.43$            & $3.67\pm0.56$   & $3.98\pm0.44$   \\
80   & $8.01\pm0.14$            & $12.71\pm0.16$           & $8.14\pm0.12$            & $8.69\pm0.13$            & $9.43\pm0.13$            & $6.93\pm0.21$            & $16.12\pm0.14$  & $16.33\pm0.14$  \\
84   & $9.51\pm0.13$            & $15.01\pm0.16$           & $9.93\pm0.13$            & $9.88\pm0.14$            & $11.04\pm0.12$           & $7.54\pm0.23$            & $19.02\pm0.13$  & $18.76\pm0.11$  \\
88   & $11.12\pm0.18$           & $17.10\pm0.23$           & $11.09\pm0.20$           & $10.95\pm0.20$           & $13.16\pm0.18$           & $8.81\pm0.21$            & $21.85\pm0.25$  & $21.11\pm0.29$  \\
92   & $12.72\pm0.19$           & $19.71\pm0.24$           & $12.54\pm0.22$           & $12.89\pm0.22$           & $14.91\pm0.19$           & $10.05\pm0.21$           & $24.93\pm0.28$  & $23.92\pm0.33$  \\
96   & $14.22\pm0.17$           & $22.15\pm0.21$           & $14.70\pm0.15$           & $14.03\pm0.26$           & $17.08\pm0.13$           & $11.52\pm0.22$           & $27.91\pm0.15$  & $26.36\pm0.16$  \\
100  & $16.84\pm0.18$           & $25.03\pm0.27$           & $16.81\pm0.18$           & $15.66\pm0.28$           & $19.15\pm0.16$           & $13.00\pm0.29$           & $31.52\pm0.17$  & $29.82\pm0.21$  \\
104  & $19.05\pm0.27$           & $27.52\pm0.36$           & $18.23\pm0.32$           & $17.42\pm0.31$           & $21.76\pm0.28$           & $15.51\pm0.31$           & $35.22\pm0.42$  & $33.82\pm0.46$  \\
108  & $21.64\pm0.30$           & $31.29\pm0.40$           & $20.30\pm0.38$           & $19.34\pm0.34$           & $24.60\pm0.36$           & $16.74\pm0.33$           & $39.26\pm0.46$  & $37.22\pm0.54$  \\
130  & $37.89\pm0.31$           & $52.80\pm0.35$           & $36.69\pm0.34$           & $31.91\pm0.35$           & $43.57\pm0.29$           & $31.76\pm0.35$           & $69.32\pm0.32$  & $60.54\pm0.27$  \\
134  & $41.87\pm0.28$           & $56.75\pm0.32$           & $40.03\pm0.31$           & $33.90\pm0.32$           & $47.47\pm0.28$           & $34.18\pm0.32$           & $74.35\pm0.33$  & $66.13\pm0.26$  \\
138  & $46.32\pm0.38$           & $62.26\pm0.40$           & $43.26\pm0.39$           & $35.91\pm0.36$           & $52.94\pm0.37$           & $38.20\pm0.39$           & $81.65\pm0.42$  & $70.47\pm0.30$  \\
142  & $49.82\pm0.37$           & $66.40\pm0.41$           & $46.90\pm0.42$           & $38.90\pm0.36$           & $57.18\pm0.39$           & $40.38\pm0.38$           & $87.93\pm0.45$  & $75.58\pm0.33$  \\
146  & $53.67\pm0.36$           & $70.06\pm0.39$           & $51.37\pm0.40$           & $41.32\pm0.35$           & $60.89\pm0.37$           & $46.23\pm0.39$           & $95.57\pm0.40$  & $79.83\pm0.32$  \\
150  & $58.07\pm0.43$           & $75.16\pm0.46$           & $54.42\pm0.47$           & $44.32\pm0.46$           & $66.20\pm0.40$           & $49.66\pm0.46$           & $101.78\pm0.46$ & $85.47\pm0.37$  \\
154  & $62.40\pm0.48$           & $79.39\pm0.57$           & $58.27\pm0.59$           & $46.49\pm0.48$           & $71.76\pm0.52$           & $53.30\pm0.51$           & $110.25\pm0.62$ & $90.22\pm0.44$  \\
158  & $67.07\pm0.55$           & $84.95\pm0.64$           & $62.41\pm0.66$           & $49.49\pm0.55$           & $77.34\pm0.60$           & $56.62\pm0.55$           & $118.45\pm0.73$ & $97.15\pm0.49$  \\
194  & $106.9\pm4.1$            & $128.8\pm4.1$            & $101.8\pm4.3$            & $72.8\pm4.1$             & $131.9\pm3.5$            & $98.8\pm4.1$             & $191.1\pm3.5$   & $158.2\pm3.8$   \\
200  & $118.3\pm3.5$            & $143.1\pm3.4$            & $105.2\pm3.7$            & $80.0\pm3.4$             & $138.1\pm3.0$            & $109.8\pm3.5$            & $213.1\pm2.9$   & $176.8\pm3.3$   \\
214  & $133.4\pm2.6$            & $164.5\pm2.7$            & $123.8\pm2.8$            & $86.3\pm2.6$             & $162.1\pm2.4$            & $124.6\pm2.7$            & $248.9\pm2.2$   & $197.0\pm2.4$   \\
220  & $143.2\pm2.6$            & $178.8\pm2.8$            & $134.1\pm2.9$            & $91.0\pm2.7$             & $179.2\pm2.3$            & $141.3\pm2.6$            & $274.0\pm2.3$   & $212.3\pm2.4$   \\
234  & $164.4\pm3.6$            & $191.4\pm3.9$            & $153.4\pm3.8$            & $103.0\pm3.4$            & $203.7\pm3.2$            & $155.5\pm4.0$            & $308.9\pm3.7$   & $236.5\pm3.2$   \\
240  & $181.1\pm4.6$            & $213.2\pm4.6$            & $167.0\pm4.9$            & $111.6\pm4.6$            & $226.8\pm4.2$            & $174.5\pm4.7$            & $338.8\pm3.9$   & $259.8\pm4.4$   \\
246  & $180.8\pm4.6$            & $225.7\pm5.1$            & $171.2\pm4.8$            & $114.1\pm3.9$            & $241.6\pm4.5$            & $178.8\pm4.9$            & $351.1\pm4.5$   & $262.7\pm3.7$   \\
252  & $196.9\pm4.1$            & $239.3\pm4.7$            & $175.6\pm4.0$            & $121.5\pm4.5$            & $255.7\pm4.2$            & $188.5\pm4.3$            & $384.9\pm4.4$   & $282.0\pm3.1$   \\
266  & $218.0\pm3.6$            & $250.1\pm3.7$            & $199.9\pm3.5$            & $129.7\pm3.4$            & $274.5\pm3.7$            & $209.8\pm3.6$            & $432.8\pm4.1$   & $304.0\pm3.0$   \\
272  & $237.8\pm5.4$            & $270.9\pm5.5$            & $213.4\pm4.6$            & $138.3\pm4.6$            & $302.1\pm4.8$            & $231.6\pm5.8$            & $470.4\pm5.2$   & $331.2\pm3.8$   \\
286  & $260.4\pm6.8$            & $288.3\pm6.2$            & $223.9\pm5.6$            & $145.5\pm5.1$            & $332.2\pm5.8$            & $244.2\pm6.3$            & $511.7\pm6.0$   & $355.3\pm4.2$   \\
292  & $271.3\pm5.6$            & $298.2\pm5.1$            & $236.5\pm5.6$            & $160.4\pm5.6$            & $351.1\pm5.1$            & $257.5\pm5.7$            & $537.7\pm6.4$   & $382.4\pm4.6$   \\
314  & $328.6\pm9.9$            & $343.0\pm7.7$            & $274.4\pm6.7$            & $173.7\pm7.3$            & $415.5\pm7.9$            & $297.3\pm7.7$            & $628.1\pm8.3$   & $439.1\pm7.4$   \\
336  & $368.9\pm9.1$            & $391.9\pm8.0$            & $304.2\pm7.5$            & $199.6\pm8.0$            & $480.2\pm8.0$            & $352.3\pm8.1$            & $710.6\pm9.0$   & $496.6\pm8.2$   \\
356  & $436\pm24$               & $434\pm17$               & $341\pm17$               & $239\pm19$               & $571\pm17$               & $399\pm19$               & $814\pm16$      & $561\pm17$      
\enddata
\tablecomments{Highlighted entries denote cases where $S_\nu / \sigma_S \le 3$, which are shown in plots as upper limits at $S_\nu + 3 \sigma_S$. The uncertainty $\sigma_S$ is only the statistical contribution -- the systematic flux calibration uncertainties are treated separately (see Sect.~\ref{sec:spec_model}).}
\end{deluxetable*}

\begin{figure*}[t!]
    \centering
    \includegraphics[width=\linewidth]{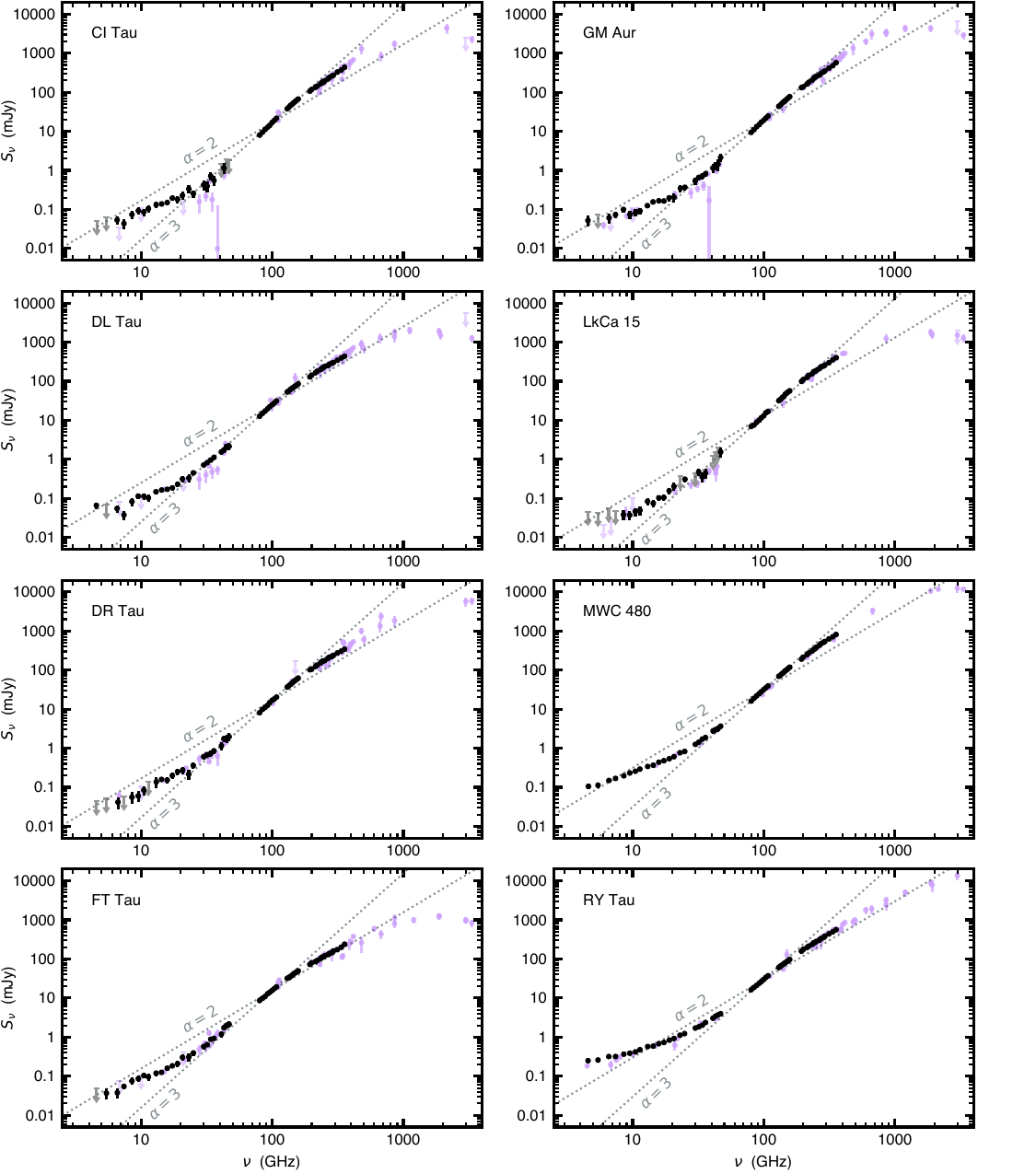}
    \caption{The microwave continuum spectra measured for each target (black points; Table \ref{table:spectra}).  Upper limits (99\%\ confidence) are marked as gray arrows.  Additional measurements from the literature (Sect.~\ref{sec:lit}) are marked in purple. The gray dotted lines show power-law spectra with fixed indices $\alpha=2$ and 3, normalized to the emission from each target at 100 GHz, as references.  }
    \label{fig:spectra_gallery}
\end{figure*}

The inferred $S_\nu$ and uncertainties (posterior medians and 68\%\ confidence intervals) for the 55 sub-bands that comprise the spectrum for each disk are listed in Table \ref{table:spectra} and are also made available as machine-readable files at \url{https://zenodo.org/records/16322048}.  The VLA-based $S_\nu$ measurements for DR Tau and RY Tau ($\nu < 50$ GHz) were corrected for minor primary beam attenuation, since the derived offsets were $\sim$10\arcsec\ from the phase centers (due to coordinate transcription errors in the observing scripts).  Figure \ref{fig:spectra_gallery} shows the measured spectra, along with previous measurements from the literature (Sect.~\ref{sec:lit}), for each target on the same scale.

\subsection{Empirical Spectrum Modeling} \label{sec:spec_model}

The microwave spectra in Figure \ref{fig:spectra_gallery} all show a distinctive pattern: a shallow spectrum contribution at low frequencies that we attribute to non-dust ``contamination'' (free-free emission); a much steeper regime at intermediate frequencies associated with an (at least partially) optically thin dust continuum; and a flattening at higher frequencies where that dust emission becomes more optically thick.  Physical models of these spectra will remain incomplete without using additional information from spatially resolved observations \citep[e.g.,][]{carrasco-gonzalez19,macias21}.  However, the dense frequency sampling presents an opportunity to characterize the detailed spectra from a more empirical perspective.

\subsubsection{Forward Modeling Framework} \label{sec:forward_modeling}

Our modeling goals were to robustly disentangle the dust and contamination contributions in the critical overlap ($\sim$cm) regime and then measure the shapes and fluxes of the dust spectra at optically thin(ner) frequencies.  To that end, we adopted a forward-modeling framework designed to sample the posterior probability distributions for a set of parameters $\theta$ that define a model $\sf{S}_{\nu}$ of the observed spectrum $S_\nu$ and uncertainties (Table \ref{table:spectra}).  The total uncertainty on a given datapoint is the combination of statistical ($\sigma_S$) and systematic contributions, where the latter is dominated by the intrinsic ambiguity of the flux calibration.  We usually think of the systematics contribution as drawn from a Gaussian, so the net variance is the quadrature sum $\sigma_\text{tot}^2 = \sigma_S^2 + (\sigma_\text{sys} S_\nu)^2$.  However, this simplification is inappropriate in our case because ``groups'' of frequencies {\it share} the same calibration factors (either by definition or because we `aligned' sub-groups based on overlaps; see Sect.~\ref{sec:obs}).  The eleven distinct calibration groups outlined in Table \ref{table:groups} are present among the 55 individual datapoints for each target.  We treated these correlated uncertainty terms with a set of nuisance parameters \{$\delta_g$\} directly in the inferences.  In practice, we assumed a Gaussian likelihood function,
\begin{equation}
    \ln {\sf p}(S_\nu | \, \theta) \propto - \frac{1}{2} \sum_g^{11} \Bigg[ \sum_j^{N_g} \left( \frac{\delta_g \, S_j - \mathsf{S}_j(\theta)}{\delta_g\,\sigma_{S_j}}\right)^{\!\! 2} \,\Bigg],
\end{equation}
where the summations are over the $N_g$ datapoints in each group and then over each group.  In essence, this means that there are 11 additional parameters (one per $\delta_g$) in the inference.  We adopted Gaussian priors for each, with means of unity and the standard deviations in Table \ref{table:groups} based on the best available estimates for each setting from each observing facility.  Marginalizing over these nuisance parameters naturally accounts for the calibration systematics in the context of the model.

\begin{deluxetable}{c | c c c c}[ht!]
\tabletypesize{\footnotesize}
\tablecaption{Shared Calibration Groups \label{table:groups}}
\tablehead{
\colhead{group $g$} & \colhead{setting(s)} & \colhead{$\nu$ range (GHz)} & \colhead{$N_g$} & \colhead{$\sigma_g$}
}
\startdata
0  & VLA C\phantom{a}     & 4 -- 8     & 4 & 0.05 \\
1  & VLA X\phantom{a}     & 8 -- 12    & 4 & 0.05 \\
2  & VLA Ku    & 12 -- 18   & 4 & 0.05 \\ 
3  & VLA K\phantom{a}     & 18 -- 26   & 4 & 0.10 \\
4  & VLA Ka    & 29 -- 37   & 4 & 0.10 \\
5  & VLA Q\phantom{a}     & 40 -- 48   & 4 & 0.10 \\
6  & NOEMA 1+2 & 78 -- 110  & 8 & 0.08 \\
7  & NOEMA 3+4 & 128 -- 160 & 8 & 0.10 \\
8  & SMA 1     & 191 -- 243 & 6 & 0.10 \\
9  & SMA 2     & 243 -- 295 & 6 & 0.10 \\
10 & SMA 3     & 309 -- 361 & 3 & 0.10 
\enddata
\end{deluxetable}

For a chosen set of priors for the model spectrum parameters $\sf{p}(\theta)$, we initialized 96 walkers using random draws from the priors and then sampled the posterior distributions (for $\theta$ and the set of $\delta_g$) using {\tt emcee} for $5\times10^6$ steps (because the autocorrelation times with so many parameters are large, $\sim$10$^4$--10$^5$ steps).  We discarded $10\times$ the autocorrelation time as burn-in.  A variety of spectrum model prescriptions were explored.  In all cases, we assume there was a linear combination of contributions from the contamination and dust continuum, $\mathsf{S}_\nu = \mathsf{S}_\nu^{\,\sf c} + \mathsf{S}_\nu^{\,\sf d}$.  The parameterizations for those contributions are outlined in the next sections.

\subsubsection{Contamination Contribution} \label{sec:contam_models}

At lower frequencies ($\lesssim$\,20 GHz), the spectra in Figure \ref{fig:spectra_gallery} are monotonic and increasing.  There are no signs in the VLA bands of peaks that would indicate a spinning dust contribution \citep{rafikov06,hoang18} or the negative spectral indices that are expected from synchrotron emission \citep{chiang96,guedel02}.  While we cannot rule out a combination of multiple mechanisms, the most straightforward interpretation is that the contamination in these targets is dominated by free-free emission, likely from jets, outflows, or winds \citep{reynolds86,pascucci12,pascucci14,anglada18}.

\out{Comparing to literature data at similar frequencies, we see no strong evidence for variability in this contribution (see Fig.~\ref{fig:spectra_gallery}).  The largest apparent discrepancies ($\sim$2--4$\sigma$) come from 2007 Green Bank Telescope (GBT) observations of CI Tau, GM Aur, and DL Tau at 28--38 GHz reported by \citet{greaves22}.  As noted by \citet{chung25}, those measurements with an early Ka band continuum instrument are systematically low (in these and other targets) compared to the VLA measurements since 2010 (see also \citealt{macias16,garufi25,zagaria25}), which are all consistent with the fluxes in Table \ref{table:spectra}.  We suggest these marginal differences are better attributed to the low SNR and challenging calibration of the  GBT data than any intrinsic variability in the targets; but certainly a more complete exploration of these (and similar) targets in the time domain would be a valuable path of future study.}

We considered two approaches for models of this free-free contribution.  The first was a broken power-law,
\begin{equation}
    \mathsf{S}_\nu^{\,\sf c} = 
        \begin{cases}
            \chi_0 S_0 \left(\frac{\nu}{\nu_0}\right)^{\!\alpha_{\rm c}} & \text{if $\nu \le \nu_{\rm c}$ ;} \\
            \chi_0 S_0 \left(\frac{\nu_{\rm c}}{\nu_0}\right)^{\!\alpha_{\rm c}} \left(\frac{\nu}{\nu_0}\right)^{\!-0.1} & \text{otherwise},
        \end{cases} 
        \label{eq:contam1}
\end{equation}
where $\chi_0$ is the fraction of the total flux $S_0$ at frequency $\nu_0$ contributed by the contamination, and $\nu_{\rm c}$ crudely mimics a transition frequency in the free-free emission spectrum from (at least partially) optically thick to thin \citep{mezger67}.  We set the reference frequency to $\nu_0 = 33$ GHz.  In the inferences, we assigned uniform priors for $\chi_0$ (0 to 1), $\alpha_\text{c}$ ($-$3 to 3), and $\log{\nu_c}$ (from 1 to 3, or equivalently 10 to 1000 GHz).  The prior used for $S_0$ was a broad Gaussian distribution, with a mean based on the measured photometry (Table \ref{table:spectra}) and a standard deviation set at 50\%\ of the mean.

The second approach considers a prescription based on a more physically realistic free-free emission model \citep[see][]{coughlan17,liu24} that smoothly transitions between optically thick ($\alpha \approx 2$) and thin ($\alpha \approx -0.1$) spectral indices,
\begin{equation}
    \mathsf{S}_\nu^{\,\sf c} = \chi_0 S_0 \left(\frac{\nu}{\nu_0}\right)^2 \Bigg(\frac{1 - \exp{[-\tau_{\rm c}(\,\nu\, / \nu_c)^{-2.1}]}}{1 - \exp{[-\tau_{\rm c}(\nu_0 / \nu_c)^{-2.1}]}}\Bigg),
    \label{eq:contam2}
\end{equation}
where $\tau_{\rm c}$ is the free-free optical depth at $\nu_c$.  We assigned a uniform prior on $\log{\tau_{\rm c}}$ from $-2$ to 2 for this prescription, and adopted the same priors as described above for Equation~(\ref{eq:contam1}) for the other parameters.

\subsubsection{Dust Contribution} \label{sec:dust_models}

An empirical prescription that describes the dust contribution, $\mathsf{S}_\nu^{\,\sf d}$, is less obvious.  Figure \ref{fig:spectra_gallery} shows that the spectra at higher frequencies cannot be explained with a single power-law: any successful model needs to flatten with increasing $\nu$.  We explored three options, which allowed us to then compare the constraints on spectral {\it curvature} for different underlying models.  

First, we considered a polynomial expansion, 
\begin{equation}
    \log{\mathsf{S}_\nu^{\,\sf d}} = \sum_{j=0}^{N_\text{poly}} a_j \Bigg[ \log{\left(\frac{\nu}{\nu_0}\right)} \Bigg]^j
    \label{eq:poly_exp}
\end{equation}
\citep{perley17}.  In this case, $a_0 = \log{(1 - \chi_0) S_0}$ is the flux contributed by dust at $\nu_0$, $a_1$ is the local spectral index ($a_1 \approx \alpha(\nu_0)$), and higher order coefficients shape the detailed curvature.  We found that a quartic expansion ($N_\text{poly} = 4$) was necessary to remove systematic residuals, which implies an additional four parameters $[a_1, a_2, a_3, a_4]$ ($a_0$ uses the same parameters that describe the contamination spectrum).  We adopted broad uniform priors for the coefficients, with ${\sf p}(a_1)$ ranging from 0 to 5 and the other terms spanning $\pm$5.

\begin{deluxetable*}{l | c c c c c c c c}[t!]
\tabletypesize{\scriptsize}
\tablecaption{Fiducial Spectrum Modeling Posterior Summaries \label{table:fid_params}}
\tablehead{
\colhead{} & \colhead{CI Tau} & \colhead{DL Tau} & \colhead{DR Tau} & \colhead{FT Tau} & \colhead{GM Aur} & \colhead{LkCa 15} & \colhead{MWC 480} & \colhead{RY Tau}
}
\startdata
$S_0$ (mJy)                      & $0.47^{+0.04}_{-0.03}$            & $0.83^{+0.05}_{-0.05}$            & $0.69^{+0.06}_{-0.05}$            & $0.75^{+0.07}_{-0.05}$           
                                 & $0.60^{+0.04}_{-0.03}$            & $0.48^{+0.04}_{-0.05}$            & $1.39^{+0.06}_{-0.05}$            & $1.88^{+0.08}_{-0.08}$ \\[2mm]
$\chi_0$                         & $0.42^{+0.07}_{-0.07}$            & $0.18^{+0.05}_{-0.04}$            & $0.44^{+0.15}_{-0.10}$            & $0.15^{+0.09}_{-0.04}$           
                                 & $0.22^{+0.05}_{-0.04}$            & $0.72^{+0.12}_{-0.41}$            & $0.25^{+0.03}_{-0.03}$            & $0.20^{+0.02}_{-0.02}$ \\[2mm]
$\log{\tau_{\rm c}}$             & $< 1.2$                           & $< 1.2$                           & $\sim{\sf p}(\log{\tau_{\rm c}})$ & $\sim{\sf p}(\log{\tau_{\rm c}})$           
                                 & $< 0.9$                           & $\sim{\sf p}(\log{\tau_{\rm c}})$ & $< 0.8$                           & $< 0.2$                \\[2mm]
$\log{(\nu_{\rm c}/\text{GHz})}$ & $\sim{\sf p}(\log{\nu_{\rm c}})$  & $\sim{\sf p}(\log{\nu_{\rm c}})$  & $\sim{\sf p}(\log{\nu_{\rm c}})$  & $\sim{\sf p}(\log{\nu_{\rm c}})$ 
                                 & $\sim{\sf p}(\log{\nu_{\rm c}})$  & $\sim{\sf p}(\log{\nu_{\rm c}})$  & $\sim{\sf p}(\log{\nu_{\rm c}})$  & $\sim{\sf p}(\log{\nu_{\rm c}})$ \\[2mm]
$\log{(\nu_{\rm d}/\text{GHz})}$ & $< 2.2$                           & $< 2.1$                           & $< 2.2$                           & $1.9^{+0.2}_{-0.2}$                          
                                 & $2.2^{+0.1}_{-0.3}$               & $2.1^{+0.2}_{-0.3}$               & $2.2^{+0.2}_{-0.3}$               & $2.1^{+0.1}_{-0.1}$    \\[2mm]
$\eta_-$                         & $4.4^{+0.4}_{-0.6}$               & $4.2^{+0.3}_{-0.5}$               & $4.5^{+0.7}_{-0.6}$               & $3.6^{+0.5}_{-0.5}$           
                                 & $3.7^{+0.6}_{-0.4}$               & $4.7^{+1.1}_{-1.1}$               & $3.7^{+0.5}_{-0.4}$               & $2.76^{+0.08}_{-0.05}$ \\[2mm]
$\eta_+$                         & $2.6^{+0.2}_{-0.2}$               & $2.3^{+0.2}_{-0.2}$               & $2.0^{+0.4}_{-0.9}$               & $2.1^{+0.3}_{-0.3}$           
                                 & $2.4^{+0.2}_{-0.3}$               & $2.3^{+0.5}_{-0.9}$               & $2.0^{+0.4}_{-1.0}$               & $2.14^{+0.06}_{-0.05}$ \\[2mm]
$\gamma$                         & $-2.0^{+0.2}_{-0.1}$\phantom{$-$} & $-2.0^{+0.1}_{-0.1}$\phantom{$-$} & $-2.2^{+0.2}_{-0.3}$\phantom{$-$} & $-2.0^{+0.2}_{-0.2}$\phantom{$-$}          
                                 & $-2.1^{+0.3}_{-0.3}$\phantom{$-$} & $-2.1^{+0.3}_{-0.3}$\phantom{$-$} & $-2.4^{+0.2}_{-0.2}$\phantom{$-$} & $-1.5^{+0.1}_{-0.2}$\phantom{$-$} 
\enddata
\tablecomments{The `best-fit' values denote the posterior medians, and the uncertainties represent 68\%\ confidence intervals.  Upper limits are taken at the 99\%\ confidence level.  The $\sim {\sf p}(X)$ notation indicates that the posteriors are consistent with the adopted priors.}  
\end{deluxetable*}

Next, we considered a piecewise power-law,
\begin{equation}
    \mathsf{S}_\nu^{\,\sf d} = F_j(\nu) \quad \text{for } f_{j-1} \le \nu < f_j
    \label{eq:segs}
\end{equation}
where the (recursive) function
\begin{equation}
    F_j(\nu) = F_{j-1}(f_j) \left(\frac{\nu}{f_{j-1}}\right)^{\alpha_j}
\end{equation}
when $j \ge 1$.  To properly normalize, we asserted 
\begin{equation}
    F_0(\nu) = (1 - \chi_0) S_0 \left(\frac{\nu}{\nu_0}\right)^{\alpha_0}
\end{equation}
and defined $f_{-1} = 0$.  At least four segments ($j \ge 3$) were required to remove systematic residuals.  In principle, this contributes seven additional parameters: the set of pivot frequencies $[f_0, f_1, f_2]$ and the segment indices $[\alpha_0, \alpha_1, \alpha_2, \alpha_3]$.  However, it was impractical to find converged MCMC chains using arbitrary pivot frequencies without imposing stringent priors.  To simplify, we experimented with several sets of fixed pivots: finding no important differences, we assigned $f_0, f_1, f_2 = 86$, 140, and 220 GHz, and inferred the remaining four indices using uniform priors from 0 to 5 in each segment.    

Finally, we considered an analogous power-law form with a continuous, frequency-dependent index, 
\begin{equation}
        \mathsf{S}_\nu^{\,\sf d} = (1 - \chi_0) S_0 \left(\frac{\nu_{\rm d}}{\nu_0}\right)^{\!\eta(\nu_0)} \! \left(\frac{\nu}{\nu_{\rm d}}\right)^{\!\eta(\nu)}, 
        \label{eq:free_index}
\end{equation}
where we specifically set
\begin{equation}
    \eta(\nu) = \eta_- + (\eta_+ - \eta_-) \left[1 + e^{-\gamma(\nu - \nu_{\rm d})}\right]^{-1},
    \label{eq:logistic_index}
\end{equation}
a sigmoid function that transitions between $\eta_-$ (at $\nu \ll \nu_{\rm d}$) and $\eta_+$ (at $\nu \gg \nu_{\rm d}$) at a rate based on  $\gamma$ (higher $\gamma$ means a sharper transition).  This model also adds four parameters, $[\eta_-, \eta_+, \nu_{\rm d}, \gamma]$.  We assigned uniform priors for $\eta_-$, $\eta_+$ ($-5$ to 10), and $\log{\nu_{\rm d}}$ ($-0.4$ to 2.6, or 40 to 400 GHz), and a Gaussian prior for $\log{\gamma}$ with a mean of $-2$ and a standard deviation of 0.5.  Note that $\eta(\nu)$ is {\it not} the spectral index $\alpha(\nu)$: since $\alpha(\nu) \equiv d\log{S_\nu}/d\log{\nu}$, the chain rule applied to Equation (\ref{eq:free_index}) implies
\begin{equation}
    \alpha(\nu) = \eta(\nu) + \frac{d \eta}{d \nu} \,\, \nu \, \log{\left(\frac{\nu}{\nu_{\rm d}}\right)}
    \label{eq:eta_to_alpha}
\end{equation}
for this kind of generalized prescription.

\section{Results} \label{sec:results}

\subsection{Spectrum Modeling Results} \label{sec:model_results}

\begin{figure}[bt!]
    \centering
    \includegraphics[width=\linewidth]{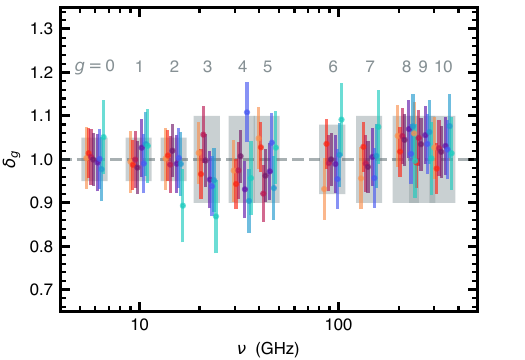}
    \caption{The inferred $\delta_g$ (68\%\ confidence intervals) as a function of group frequency (each target is offset by a small amount around the central group frequency for clarity): see Table \ref{table:groups}.  The colors correspond to the different targets, and match the annotations in other figures.  The adopted priors (68\%\ confidence intervals, corresponding to the standard uncertainties adopted for the absolute flux calibration) are denoted by shaded gray boxes.}
    \label{fig:nuisances}
\end{figure}

\begin{figure*}[ht!]
    \centering
    \includegraphics[width=\linewidth]{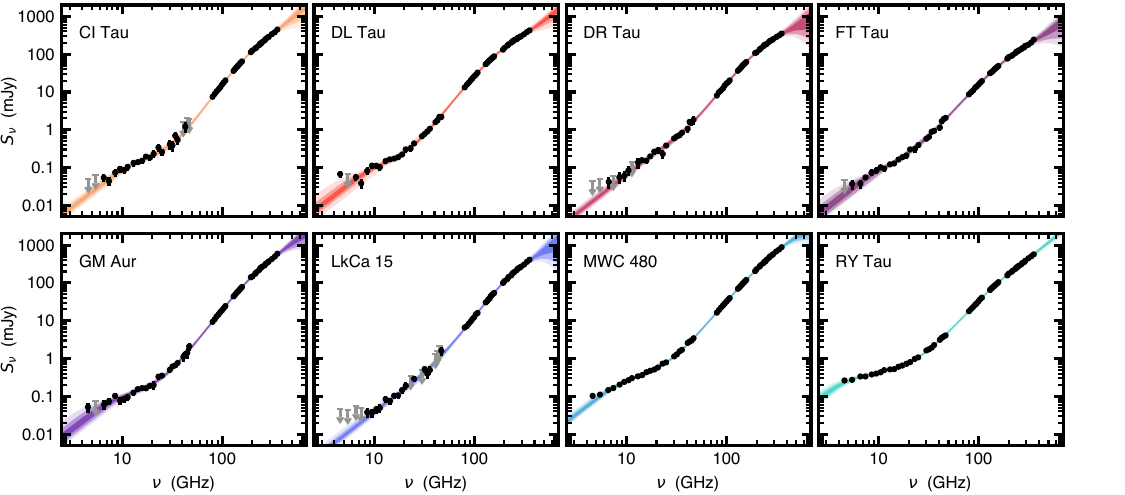}
    \caption{Comparisons of the observed spectra (as in Fig.~\ref{fig:spectra_gallery}) with models reconstructed from the inferred posterior distributions for the logistic index prescription (Eq.~\ref{eq:free_index}).  The dark to light color shadings correspond to the 68, 95, and 99\%\ confidence intervals.  (For reference, an analogous figure in the Appendix highlights the decomposition of the model into the contamination and dust emission components.)}
    \label{fig:fits_gallery}
\end{figure*}

\begin{figure*}[ht!]
    \centering
    \includegraphics[width=\linewidth]{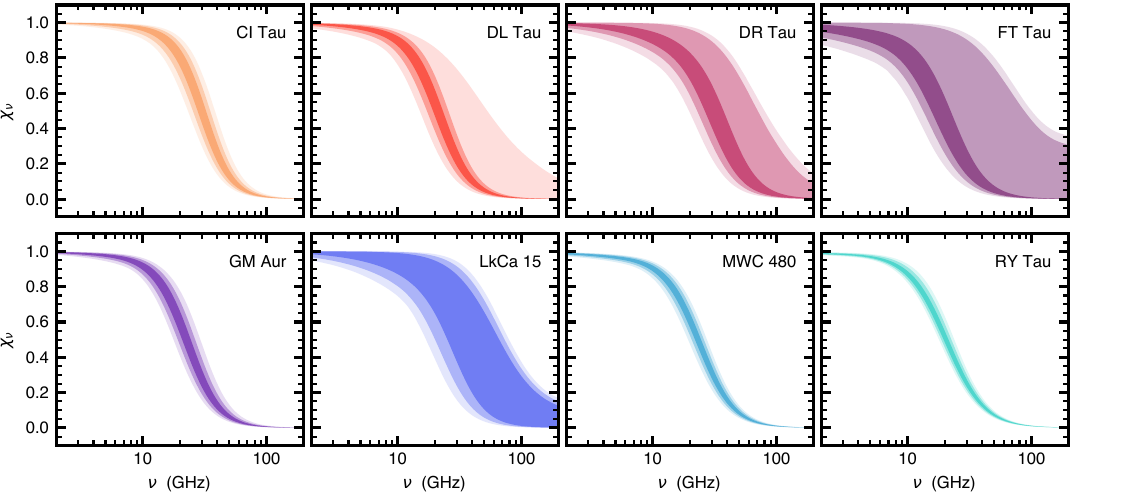}
    \caption{The marginalized posterior distributions of $\chi_\nu$, the fraction of the total flux contributed by the (free-free) contamination, as a function of frequency (i.e., $\chi_\nu \equiv \mathsf{S}_\nu^{\, \sf c} / (\mathsf{S}_\nu^{\, \sf c} + \mathsf{S}_\nu^{\, \sf d})$).  The dark to light color shadings correspond to the 68, 95, and 99\%\ confidence intervals.}
    \label{fig:chinu}
\end{figure*}

We found excellent fits to the observed spectra from all combinations of the model prescriptions outlined above.  There are no important differences between the conclusions drawn from the different prescriptions, so for the sake of clarity we will adopt the more physically motivated contamination model (Eq.~\ref{eq:contam2}) and the sigmoid index dust model (Eq.~\ref{eq:free_index}); the latter is a better-performing selection for either contamination model ($\Delta \chi^2 \approx 0.5$--7 compared to the other options; note that all prescriptions have identical degrees of freedom).  Some further discussion of the alternative model combinations is provided in the Appendix.  Summaries of the posteriors for this ``fiducial'' model prescription are compiled in Table \ref{table:fid_params}.  Figure \ref{fig:nuisances} demonstrates that the posterior distributions inferred for the nuisance parameters $\delta_g$ are consistent with the priors based on the intrinsic flux calibration uncertainties.  We note that simplified fits where the calibration covariances between data groups are ignored (each datapoint has an independent systematic uncertainty added in quadrature to the formal statistical uncertainty) provide indistinguishable posterior behaviors (i.e., the accuracy is unaffected).  However, the method outlined in Section~\ref{sec:forward_modeling} offers a slight improvement in precision because it takes advantage of the available information on the intra-group spectral variations.

Figure \ref{fig:fits_gallery} compares the measured spectra with reconstructed models based on the posterior samples.  No constraints are available for the contamination turnover frequencies $\nu_{\rm c}$, and only upper limits could be derived for the free-free optical depths $\tau_{\rm c}$ ($\nu_{\rm c}$ and $\tau_{\rm c}$ are highly correlated over those ranges) and in some cases the dust turnover frequencies $\nu_d$ (entries in Table \ref{table:fid_params} denote 99\%\ confidence intervals), but those ambiguities contribute to the marginalized posteriors for other parameters.

\begin{figure*}[ht!]
    \centering
    \includegraphics[width=\linewidth]{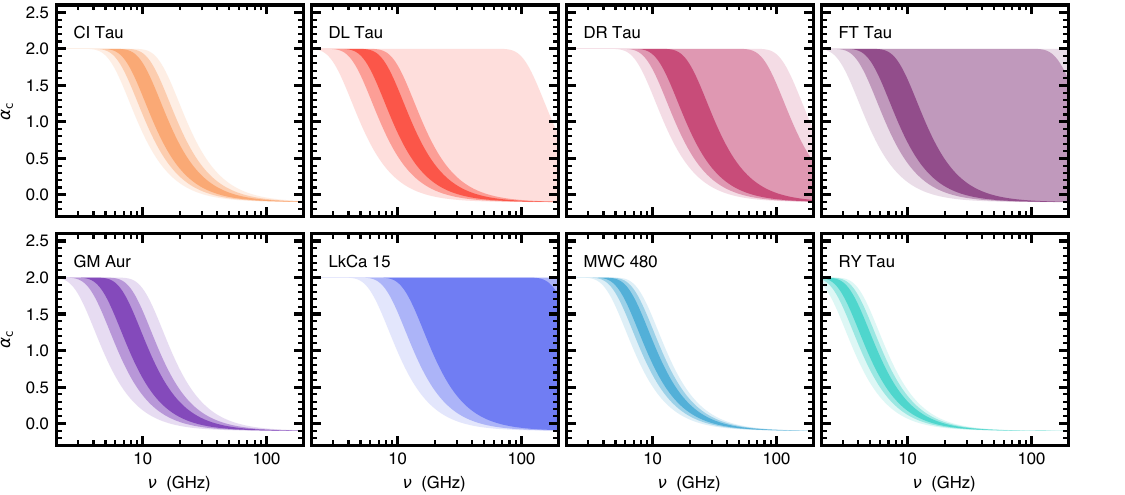}
    \caption{The inferred frequency variations of the contamination (free-free) spectral index for each target.  The shaded regions trace the posterior distributions at 68, 95, and 99\%\ confidence intervals ($\sim$1, 2, and 3$\sigma$) from darker to lighter.}
    \label{fig:alphac_gallery}
\end{figure*}

Figure \ref{fig:chinu} shows the (marginalized) posterior distributions for the fraction of the total emission contributed by the (free-free) contamination as a function of frequency, $\chi_\nu$ (note the parameter $\chi_0$ is defined at $\nu_0 = 33$ GHz).  In all cases, the free-free emission dominates the spectrum below $\sim$20 GHz (K band), but contamination fractions are low enough at $\gtrsim$\,30 GHz (Ka and Q bands) to permit meaningful constraints on the local dust spectra (as described below).  The ability to leverage those frequencies with mixed contributions is significantly aided by the dense spectral sampling of the data.  

\begin{figure}[ht!]
    \centering
    \includegraphics[width=\linewidth]{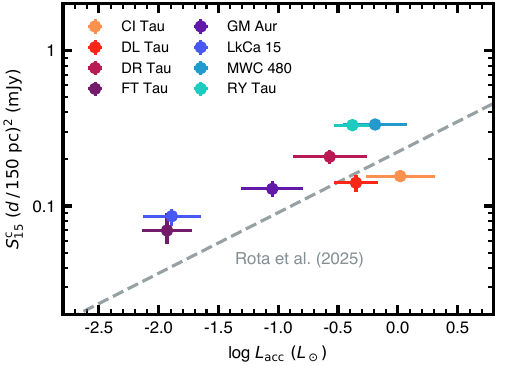}
    \caption{\out{The inferred 15 GHz fluxes (scaled for distance) contributed by contamination compared to the stellar accretion rates derived by \citet{gangi22}.  These measurements are consistent with the correlation derived by \citet{rota25} for a different (but partially overlapping) sample.}}
    \label{fig:rota}
\end{figure}

Figure \ref{fig:alphac_gallery} shows the posterior distributions for the local spectral {\it curvature} of the contamination contribution, $\alpha_{\rm c}(\nu) \equiv d \log{\mathsf{S}_\nu^{\, \sf c}} / d\log{\nu}$.  Given the adopted model prescription, we see the characteristic behavior where the spectrum flattens from an optically thick $\alpha_{\rm c} \approx 2$ at low frequencies to the optically thin limit $\alpha_{\rm c} \approx -0.1$ by $\sim$20--60 GHz.  But much of the {\it direct} information available from the data comes at $\nu \lesssim 20$ GHz (where $\chi_\nu \gtrsim 0.5$), in the transition between these two bounds.  The result is a contamination spectrum with an apparent $\alpha_{\rm c} \approx 0.5$--1, consistent with most previous measurements in the literature \citep{ubach12,ubach17,garufi25,chung25}.  \out{The normalizations of these contamination spectra are also in line with similar targets in the literature.  Figure \ref{fig:rota} compares the inferred 15 GHz fluxes (normalized by the distances from \citealt{gaia_dr3}) with the stellar accretion luminosities estimated by \citet{gangi22} and finds behavior consistent with the scaling relation found by \citet{rota25}.}  This general behavior suggests that the mass-loss rates in the ionized jets/outflows are indeed connected to the accretion process \citep[e.g.,][]{hartigan95,anglada18,rota25}.

\begin{figure*}[ht!]
    \centering
    \includegraphics[width=\linewidth]{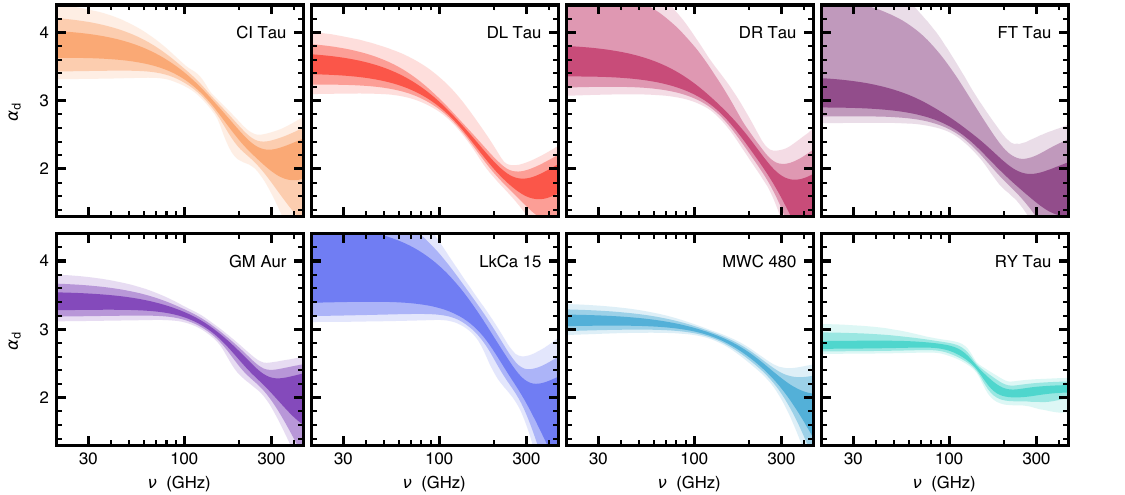}
    \caption{The inferred frequency variations of the dust spectral index -- the spectral curvature -- for each target.  The shaded regions trace the posterior distributions at 68, 95, and 99\%\ confidence intervals ($\sim$1, 2, and 3$\sigma$) from darker to lighter.}
    \label{fig:alphad_gallery}
\end{figure*}

Figure \ref{fig:alphad_gallery} shows the inferred spectral curvature of the dust emission, the frequency variation of the spectral index $\alpha_{\rm d}(\nu) \equiv d \log{\mathsf{S}_\nu^{\, \sf d}} / d\log{\nu}$ (or, in this case, see Eq.~\ref{eq:eta_to_alpha}), for each target.  As we noted in Section \ref{sec:analysis} from a visual inspection of the observations (Fig.~\ref{fig:spectra_gallery}), all of the spectra flatten ($\alpha_{\rm d}$ decreases) with increasing $\nu$.  At the higher frequency ($\sim$mm; $\gtrsim$\,200 GHz) bands that are commonly used to study disks, we find $\alpha_{\rm d} \approx 2$.  This is consistent with the mean spectral index at these frequencies for much larger and more diverse disk samples \citep[e.g.,][]{ricci10b,ricci10a,andrews20,tazzari21,chung24}, and suggests that a significant fraction of the emission at those frequencies is likely polluted by high optical depths.  At lower, optically thinner, frequencies ($\sim$cm; $\lesssim$\,100 GHz), the spectra become much steeper: we find $\alpha_{\rm d} \approx 2.8$--4.0 at 43 GHz for these targets.  \citet{garufi25} reached a similar conclusion with coarser spectral sampling, but a larger sample.  Presumably, the indices at lower frequencies better reflect the true local shape of the opacity spectra (see below).  Measurements of spectral curvature with the precisions shown in Figure \ref{fig:alphad_gallery} are available because of the broad and dense spectral sampling across the microwave bands.  

\begin{figure}[t!]
    \centering
    \includegraphics[width=\linewidth]{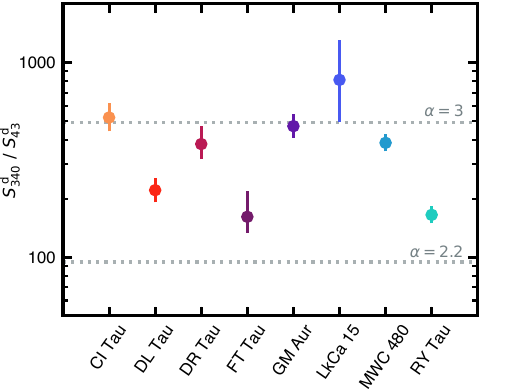}
    \caption{The ratios of the inferred dust fluxes at 340 and 43 GHz (error bars mark the 68\%\ uncertainties).  The colors are matched to the previous plots for target identification.  The dotted lines show the ratios that correspond to spectral index values of 3 and 2.2, the latter being the mean observed at $\nu \gtrsim \, 200$ GHz for large disk samples \citep{tazzari21,chung24}.}
    \label{fig:mm_cm}
\end{figure}

Because the dust spectra have different shapes, owing to variations in the underlying opacities and/or the optical depth and temperature variations, the standard task of converting a microwave continuum flux into a dust disk mass then depends on the selected observing frequency.  We explore this in the following section, but first wanted to emphasize that even studies focused only on comparing fluxes are affected (i.e., most demographic studies).  Figure \ref{fig:mm_cm} illustrates the point, comparing the ratios of 340 to 43 GHz fluxes from the dust spectra for these targets.  The key takeaway is that the ratio ranges by a factor of $\sim$5.  This may seem obvious, but the implications can be profound.  Studies that are based on the mm-band continuum fluxes could be misleading in their conclusions about the demographic behaviors of disk masses: swapping in their cm-band equivalents could in principle lead to different scaling relations (and scatter).

\subsection{Dust Mass Estimates} \label{sec:Mdust}

With these more precise characterizations of dust emission at lower microwave frequencies, we revisited some simple estimates of dust disk masses $M_{\rm d}$.  We assumed that the emission around frequency $\nu_{\rm thin}$ is optically thin.  Then, the local spectral indices $\alpha_{\rm d}(\nu_{\rm thin})$ that we inferred (Fig.~\ref{fig:alphad_gallery}) can be related to the opacity indices $\beta(\nu_{\rm thin}) \approx \alpha_{\rm d}(\nu_{\rm thin}) - 2$.  Using model absorption opacities $\kappa({\nu_{\rm thin}})$ that correspond to those $\beta(\nu_{\rm thin})$, the inferred flux densities $\mathsf{S}^{\sf d}({\nu_{\rm thin}})$ and some standard assumptions can be used to estimate $M_{\rm d}$.   

First, we defined a dust opacity index model $\beta(\theta) = d\log{\kappa_\nu}/d\log{\nu}$, characterized by the properties of the dust population.  We distilled these to four parameters $\theta = [p_{\rm d}, a_{\rm max}, f_{\rm fill}, f_{\rm AC}]$: two define a power-law particle size ($a$) distribution, including the index $p_{\rm d}$ (where $n(a) \propto a^{-p_{\rm d}}$) and maximum size $a_{\rm max}$ (the minimum size is negligible and fixed at 0.1 $\mu$m); another sets the volume filling factor $f_{\rm fill}$ of the particles (i.e., $1 - $porosity); and one more represents a compositional toggle $f_{\rm AC}$ \citep[see][]{ueda25}.  We adopted a modified DSHARP composition and optical constants \citep[see][]{dsharp5}, permitting adjustments to the fraction of the total carbonaceous material in amorphous form (with optical constants from \citealt{zubko96}) using the parameter $f_{\rm AC}$ (the remainder is in refractory organics; \citealt{henning96}).  For example, $f_{\rm AC} = 0$ is equivalent to the default DSHARP model \citep{dsharp5}, while $f_{\rm AC} \approx 0.4$ and 0.9 provide opacities similar to the \citet[][DIANA]{woitke16} and \citet{ricci10a} models, respectively.  We used the {\tt optool} package \citep{dominik2021optool}, deploying the \citet{bruggeman35} mixing rule and the distribution of hollow spheres approach \citep{min05}, to calculate opacity spectra on a pre-tabulated grid that spans $p_d \in [2, 4]$, $f_{\rm ac} \in [0, 1]$, and $f_{\rm fill} \in [0.001, 1]$ in 20, 10, and 20 linearly-spaced samples, and $a_{\rm max} \in [10, 10^7]$ $\mu$m with 60 logarithmically-spaced samples.  A four-dimensional linear interpolation over that grid was used to define the $\beta(\theta)$ models for any arbitrary point in the domain.

\begin{figure*}[ht!]
    \centering
    \includegraphics[width=\linewidth]{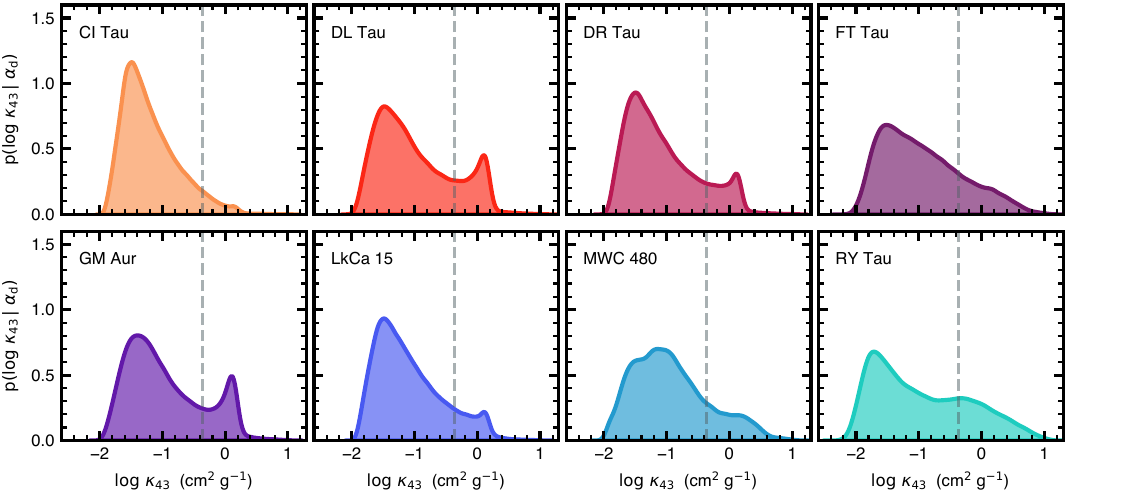}
    \caption{Posterior distributions for the 43 GHz absorption opacities, marginalized over the parameters that described the disk-averaged dust properties, and based on the constraints from the measured local dust spectral indices (see Fig.~\ref{fig:alphad_gallery}).  Gray dashed lines mark the canonical \citet{beckwith90} opacity extrapolated to 43 GHz for reference.}
    \label{fig:opacities}
\end{figure*}

\begin{figure*}[ht!]
    \centering
    \includegraphics[width=\linewidth]{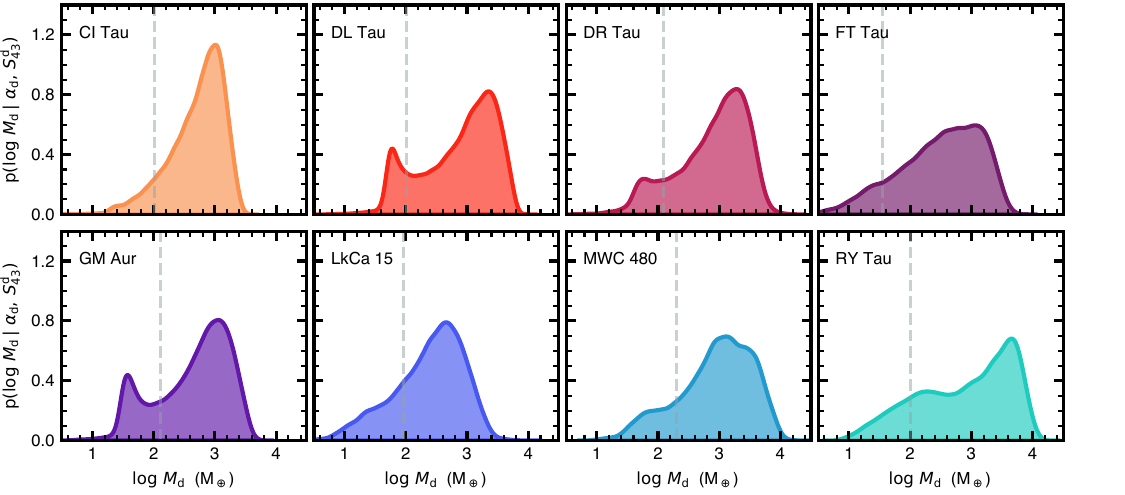}
    \caption{Constraints on the dust masses for the target disks, based on the opacities in Fig.~\ref{fig:opacities} and the 43 GHz dust fluxes ($\mathsf{S}_{43}^{\, \sf d}$) derived in the previous section.  The gray dashed lines show the mean mass that would be obtained using instead the 340 GHz fluxes and the \citet{beckwith90} opacity prescription, representative of the standard estimates in the literature.}
    \label{fig:mass}
\end{figure*}

Next, we defined a numerical objective function from the posterior samples of $\alpha_{\rm d}(\nu_{\rm thin}) - 2$ using a kernel density estimator, assuming that the Rayleigh-Jeans limit is appropriate at $\nu_{\rm thin}$ \out{(this is the case for disk-averaged temperatures $\gtrsim 20$ K for $\nu_{\rm thin} \lesssim 100$ GHz)}.  This function defines a likelihood of $\theta$ for a model opacity index $\beta(\theta)$ (at $\nu_{\rm thin}$) conditioned on $\alpha_{\rm d}(\nu_{\rm thin})$.  Adopting uniform priors across the model opacity grid domain, we sampled the posterior distributions of $\theta$ with {\tt emcee} using 64 walkers and $3\times10^5$ steps (with a burn-in phase discarded based on the autocorrelation times, $\lesssim$\,1000 steps).  For each $\theta$ posterior sample, we also collated the corresponding model absorption opacity $\kappa(\nu_{\rm thin})$.

Not surprisingly, we did not find significant constraints on $f_{\rm AC}$, $f_{\rm fill}$, or $p_{\rm d}$; and while the $a_{\rm max}$ posteriors are peaked, they have rather broad distributions (a visualization of the posteriors is provided in the Appendix).  In any case, it is not clear how much to read into these detailed grain properties with the limited information available.  Of more immediate interest are the corresponding constraints on the absorption opacities $\kappa(\nu_{\rm thin})$.  To focus, we selected $\nu_{\rm thin} = 43$ GHz because it is accessible (a standard frequency for VLA Q or ALMA Band 1 observations), likely to be optically thin, and less affected by any lingering ambiguities associated with disentangling free-free contamination (see Fig.~\ref{fig:chinu}).  The derived constraints on $\kappa_{43}$ are shown together in Figure \ref{fig:opacities}.  Despite the limited information available on the dust properties, the 43 GHz opacities are clearly peaked at $\sim$0.03--0.1 cm$^2$ g$^{-1}$ for all targets, with 68\%\ uncertainties ranging $\sim$0.4 dex (a factor of $\sim$2--3) around those peaks and with broad asymmetric tails up to higher values (about half the targets have smaller, secondary peaks that correspond to bimodal $a_{\rm max} f_{\rm fill}$ distributions; see the Appendix).   

These opacity distributions were then used to estimate corresponding distributions of the dust masses using the standard optically thin and isothermal approximation,
\begin{equation}
    M_{\rm d} = \left.\frac{S_\nu^{\rm d} \, d^2}{\kappa_\nu B_\nu(T_{\rm d})} \right\rvert_{\nu = \nu_{\rm thin}},
\label{eq:mass_eq}
\end{equation}
where $B_\nu$ denotes the Planck function and $T_{\rm d}$ is a disk-averaged temperature.  We compiled samples of $M_{\rm d}$ using draws from the $\kappa_\nu$ distributions, the posteriors for the dust emission fluxes at 43 GHz, distances from Gaia DR3 \citep{gaia_dr3}, and a fixed $T_{\rm d} = 20$ K.  The resulting dust mass constraints are shown in Figure \ref{fig:mass}. The dust mass estimates for these targets peak at $M_{\rm d} \gtrsim 1000 \, M_\oplus$, with an effective uncertainty of $\sim$0.5 dex and broad, extended tails (particularly to lower values), both inherited from the opacity constraints.  

The most probable dust masses are $\sim$10$\times$ higher than would be derived from the canonical approach that assumes the (sub)mm band emission is optically thin and has an opacity extrapolated from the \citet{beckwith90} approximation, where $\kappa_\nu \approx 10 (\nu / \text{1000 GHz})$ cm$^2$ g$^{-1}$ (i.e., with $\beta = 1$).  For reference, those values that would be derived from the 340 GHz fluxes ($\mathsf{S}_{340}^{\,\rm d}$) are marked as vertical dashed lines in Figure \ref{fig:mass}.  Some of that mass discrepancy is related to high optical depths, but much of it is because the \citet{beckwith90} opacity prescription was based on far-infrared measurements in the interstellar medium \citep{hildebrand83}, where the normalization was not modified in a physically motivated way that is consistent with the assumed lower $\beta$.

\section{Discussion} \label{sec:discussion}

We measured the detailed broadband microwave spectra from a subset of bright protoplanetary disks in Taurus and demonstrated that their thermal dust continuum emission exhibits significant spectral curvature, with local spectral indices dropping from $\alpha_{\rm d} \approx 3$--4 at the $\sim$cm-bands ($\lesssim$\,100 GHz) to $\approx 2$ in the (sub)mm regime ($\gtrsim$\,200 GHz).  This behavior suggests that the higher frequency emission -- currently used in many studies as a diagnostic of the dust mass -- is significantly polluted by high optical depths, and that the $\sim$cm-band emission is therefore a more accurate tracer of the true opacity spectrum shape ($\beta$) and dust mass in analogous disks.  

The precision of such measurements depends on access to high quality data over a broad frequency range with sufficiently dense spectral sampling.  Specifically, coverage in three frequency regimes are critical: (1) across the transition from thermal dust- to (free-free) contamination-dominated emission ($\sim$5--50 GHz), to disentangle and probe the dust spectrum shape at the optically thinnest frequencies (VLA data); (2) the turnover at $\sim$90--150 GHz, where the dust spectrum shape changes significantly due to strong variations in the optical depths (NOEMA data); and (3) an anchoring at high frequencies ($\gtrsim$\,200 GHz) to gauge where the emission is optically thick (SMA data).  Sacrificing sensitivity or using coarser spectral sampling will reduce precision \citep[e.g.,][]{chung25,garufi25}, but lacking coverage in any of those frequency regimes will compromise accuracy.

The steep spectral indices we measured at lower (optically thinner) frequencies imply lower absorption opacities than would be assumed from an extrapolation of the \citet{beckwith90} prescription.  The net results are dust masses $\sim$10$\times$ higher than would be inferred from the (optically thicker) (sub)mm fluxes. Previous studies have also suggested and found various lines of evidence supporting similar underestimates of dust mass attributed to high optical depths \citep[e.g.,][]{ribas2020modeling, macias21, sierra21, xin23, viscardi2025dust,godines25}.  While these are simplified calculations, they suggest that the mass budget available for planet formation may be substantially larger than is often appreciated.  If such underestimates, caused by high optical depths at easily accessible frequencies, are common, then this effect alone can explain the ``missing mass'' problem in disks with respect to the exoplanet population \citep{manara18}. 

However, there are important caveats to these conclusions.  One caution is related to target selection: the disks observed in this study represent only a small and biased sample, selected because they were large and bright at $\gtrsim$\,200 GHz.  Moreover, high resolution observations demonstrate that most (perhaps all) of them exhibit prominent substructures \citep{long18,clarke18,jennings22}, and therefore may be prone to high optical depths due to the associated dust concentrations (e.g., \citealt{ricci12}; but see the argument from \citealt{garufi25} that more compact disks without known substructures appear to be {\it more} optically thick).  In any case, it remains unclear how much these results can be generalized to the broader disk population.  

Beyond the sample, there are naturally some caveats associated with our analysis choices.  Even though the model we used for the spectral decompositions is more complex than is typically adopted, it may still be too simple.  For example, we assumed the contamination spectrum is dominated by a single emission mechanism, but if instead it is a combination of multiple mechanisms it is plausible that the inferred dust spectral indices could be biased.  Likewise, the dust spectrum itself could be more complicated.  For example, an additional population of cold, larger solids might preferentially contribute to the $\sim$cm-band spectrum with a flatter spectral index \citep[e.g.,][]{wilner05,hashimoto22,liu24_dmtau} and be difficult to disentangle from the contamination (though, notably, this would imply even higher $M_{\rm d}$).  These kinds of model mis-specifications are challenging to diagnose with unresolved spectra alone.

Looking ahead, these results based on the continuum spectra represent a foundation for several next steps in improving our understanding of disk solids.  One approach of particular interest is to combine these densely sampled, broadband spectra with spatially resolved measurements across a more limited, sparse spectral range \citep[e.g.,][]{carrasco-gonzalez19,macias21,ueda25} to significantly improve the constraints on the distributions of key dust properties and physical parameters.  Another future focus will be to expand the sample of targets with similar data, to examine the relationship between optically thin emission and other key disk demographic properties that have already been studied in optically thick, (sub)mm emission -- particularly stellar host mass, age, accretion rate, and disk environment. While there is growing interest in studying the cm-band emission from disks \citep{chung25,garufi25}, the current samples are still relatively small and biased.  Establishing population-level trends will be essential for preparing future large-scale resolved studies of cm-band dust emission with the ngVLA.

\acknowledgments 
We are grateful to Mark Gurwell for his advice on designing the SMA spectral setups and flux calibration strategy; Michael Bremer for his patient assistance with NOEMA data calibration; Melanie Krips and Kirsty Butler for their help with the NOEMA observations; and Scott Paine for generating the {\tt am} models in Figure \ref{fig:spectral_settings}.  

This work was based on observations with the National Radio Astronomy Observatory's Jansky Very Large Array (VLA), from project {\tt 22A--179}.  The National Radio Astronomy Observatory is a facility of the National Science Foundation operated under cooperative agreement by Associated Universities, Inc.  This work is also based on observations carried out under project number {\tt W22AX} with the IRAM NOEMA Interferometer.  IRAM is supported by INSU/CNRS (France), MPG (Germany), and IGN (Spain).  This work was also based on observations with the Submillimeter Array, as part of project {\tt 2022B--S015}.  The Submillimeter Array is a joint project between the Smithsonian Astrophysical Observatory and the Academia Sinica Institute of Astronomy and Astrophysics, and is funded by the Smithsonian Institution and the Academia Sinica.

CC-G acknowledges support from UNAM DGAPA PAPIIT grant IG101224 and from CONAHCyT Ciencia de Frontera project ID 86372. TB acknowledges funding from the European Union under the European Union's Horizon Europe Research and Innovation Programme 101124282 (EARLYBIRD) and funding by the Deutsche Forschungsgemeinschaft (DFG, German Research Foundation) under Germany's Excellence Strategy EXC-2094 - 390783311. Views and opinions expressed are, however, those of the authors only and do not necessarily reflect those of the European Union or the European Research Council. Neither the European Union nor the granting authority can be held responsible for them. \\

\noindent \textsf{\textbf{Software:}} 
{\tt astropy} \citep{astropy, astropy_old, astropy_oldest},
{\tt CASA} \citep{CASA22}, 
{\tt cmasher} \citep{cmasher},
{\tt corner} \citep{corner},
{\tt emcee} \citep{foreman-mackey13},
{\tt matplotlib} \citep{matplotlib},
{\tt numpy} \citep{numpy}, 
{\tt optool} \citep{dominik2021optool},
{\tt scipy} \citep{scipy},
{\tt seaborn} \citep{seaborn} \\

\noindent \textsf{\textbf{Facilities:}}
Jansky Very Large Array (VLA), IRAM Northern Extended Millimeter Array (NOEMA), Submillimeter Array (SMA).

\bibliography{main}{}

\begin{thebibliography}{}
\expandafter\ifx\csname natexlab\endcsname\relax\def\natexlab#1{#1}\fi
\providecommand{\url}[1]{\href{#1}{#1}}
\providecommand{\dodoi}[1]{doi:~\href{http://doi.org/#1}{\nolinkurl{#1}}}
\providecommand{\doeprint}[1]{\href{http://ascl.net/#1}{\nolinkurl{http://ascl.net/#1}}}
\providecommand{\doarXiv}[1]{\href{https://arxiv.org/abs/#1}{\nolinkurl{https://arxiv.org/abs/#1}}}

\bibitem[{{Adams} {et~al.}(1990){Adams}, {Emerson}, \& {Fuller}}]{adams90}
{Adams}, F., {Emerson}, J.~P., \& {Fuller}, G.~A. 1990, \apj, 357, 606, \dodoi{10.1086/168949}

\bibitem[{{Altenhoff} {et~al.}(1994){Altenhoff}, {Thum}, \& {Wendker}}]{altenhoff94}
{Altenhoff}, W., {Thum}, C., \& {Wendker}, H.~J. 1994, \aap, 281, 161

\bibitem[{{Andrews} {et~al.}(2024){Andrews}, {Teague}, {Wirth}, {Huang}, \& {Zhu}}]{andrews24}
{Andrews}, S., {Teague}, R., {Wirth}, C.~P., {Huang}, J., \& {Zhu}, Z. 2024, \apj, 970, 153, \dodoi{10.3847/1538-4357/ad5285}

\bibitem[{{Andrews} \& {Williams}(2005)}]{aw05}
{Andrews}, S., \& {Williams}, J. 2005, \apj, 631, 1134, \dodoi{10.1086/432712}

\bibitem[{{Andrews} \& {Williams}(2007)}]{aw07a}
---. 2007, \apj, 659, 705, \dodoi{10.1086/511741}

\bibitem[{{Andrews} {et~al.}(2011){Andrews}, {Wilner}, {Espaillat}, {Hughes}, {Dullemond}, {McClure}, {Qi}, \& {Brown}}]{andrews11}
{Andrews}, S., {Wilner}, D.~J., {Espaillat}, C., {et~al.} 2011, \apj, 732, 42, \dodoi{10.1088/0004-637X/732/1/42}

\bibitem[{{Andrews}(2020)}]{andrews20}
{Andrews}, S.~M. 2020, \araa, 58, 483, \dodoi{10.1146/annurev-astro-031220-010302}

\bibitem[{{Anglada} {et~al.}(2018){Anglada}, {Rodr{\'\i}guez}, \& {Carrasco-Gonz{\'a}lez}}]{anglada18}
{Anglada}, G., {Rodr{\'\i}guez}, L.~F., \& {Carrasco-Gonz{\'a}lez}, C. 2018, \aapr, 26, 3, \dodoi{10.1007/s00159-018-0107-z}

\bibitem[{{Astropy Collaboration} {et~al.}(2013){Astropy Collaboration}, {Robitaille}, {Tollerud}, {Greenfield}, {Droettboom}, {Bray}, {Aldcroft}, {Davis}, {Ginsburg}, {Price-Whelan}, {Kerzendorf}, {Conley}, {Crighton}, {Barbary}, {Muna}, {Ferguson}, {Grollier}, {Parikh}, {Nair}, {Unther}, {Deil}, {Woillez}, {Conseil}, {Kramer}, {Turner}, {Singer}, {Fox}, {Weaver}, {Zabalza}, {Edwards}, {Azalee Bostroem}, {Burke}, {Casey}, {Crawford}, {Dencheva}, {Ely}, {Jenness}, {Labrie}, {Lim}, {Pierfederici}, {Pontzen}, {Ptak}, {Refsdal}, {Servillat}, \& {Streicher}}]{astropy_oldest}
{Astropy Collaboration}, {Robitaille}, T.~P., {Tollerud}, E.~J., {et~al.} 2013, \aap, 558, A33, \dodoi{10.1051/0004-6361/201322068}

\bibitem[{{Astropy Collaboration} {et~al.}(2018){Astropy Collaboration}, {Price-Whelan}, {Sip{\H o}cz}, {G{\"u}nther}, {Lim}, {Crawford}, {Conseil}, {Shupe}, {Craig}, {Dencheva}, {Ginsburg}, {VanderPlas}, {Bradley}, {P{\'e}rez-Su{\'a}rez}, {de Val-Borro}, {Aldcroft}, {Cruz}, {Robitaille}, {Tollerud}, {Ardelean}, {Babej}, {Bach}, {Bachetti}, {Bakanov}, {Bamford}, {Barentsen}, {Barmby}, {Baumbach}, {Berry}, {Biscani}, {Boquien}, {Bostroem}, {Bouma}, {Brammer}, {Bray}, {Breytenbach}, {Buddelmeijer}, {Burke}, {Calderone}, {Cano Rodr{\'{\i}}guez}, {Cara}, {Cardoso}, {Cheedella}, {Copin}, {Corrales}, {Crichton}, {D'Avella}, {Deil}, {Depagne}, {Dietrich}, {Donath}, {Droettboom}, {Earl}, {Erben}, {Fabbro}, {Ferreira}, {Finethy}, {Fox}, {Garrison}, {Gibbons}, {Goldstein}, {Gommers}, {Greco}, {Greenfield}, {Groener}, {Grollier}, {Hagen}, {Hirst}, {Homeier}, {Horton}, {Hosseinzadeh}, {Hu}, {Hunkeler}, {Ivezi{\'c}}, {Jain}, {Jenness}, {Kanarek}, {Kendrew}, {Kern}, {Kerzendorf}, {Khvalko}, {King}, {Kirkby}, {Kulkarni},
  {Kumar}, {Lee}, {Lenz}, {Littlefair}, {Ma}, {Macleod}, {Mastropietro}, {McCully}, {Montagnac}, {Morris}, {Mueller}, {Mumford}, {Muna}, {Murphy}, {Nelson}, {Nguyen}, {Ninan}, {N{\"o}the}, {Ogaz}, {Oh}, {Parejko}, {Parley}, {Pascual}, {Patil}, {Patil}, {Plunkett}, {Prochaska}, {Rastogi}, {Reddy Janga}, {Sabater}, {Sakurikar}, {Seifert}, {Sherbert}, {Sherwood-Taylor}, {Shih}, {Sick}, {Silbiger}, {Singanamalla}, {Singer}, {Sladen}, {Sooley}, {Sornarajah}, {Streicher}, {Teuben}, {Thomas}, {Tremblay}, {Turner}, {Terr{\'o}n}, {van Kerkwijk}, {de la Vega}, {Watkins}, {Weaver}, {Whitmore}, {Woillez}, {Zabalza}, \& {Astropy Contributors}}]{astropy_old}
{Astropy Collaboration}, {Price-Whelan}, A.~M., {Sip{\H o}cz}, B.~M., {et~al.} 2018, \aj, 156, 123, \dodoi{10.3847/1538-3881/aabc4f}

\bibitem[{{Astropy Collaboration} {et~al.}(2022){Astropy Collaboration}, {Price-Whelan}, {Lim}, {Earl}, {Starkman}, {Bradley}, {Shupe}, {Patil}, {Corrales}, {Brasseur}, {N{\"o}the}, {Donath}, {Tollerud}, {Morris}, {Ginsburg}, {Vaher}, {Weaver}, {Tocknell}, {Jamieson}, {van Kerkwijk}, {Robitaille}, {Merry}, {Bachetti}, {G{\"u}nther}, {Aldcroft}, {Alvarado-Montes}, {Archibald}, {B{\'o}di}, {Bapat}, {Barentsen}, {Baz{\'a}n}, {Biswas}, {Boquien}, {Burke}, {Cara}, {Cara}, {Conroy}, {Conseil}, {Craig}, {Cross}, {Cruz}, {D'Eugenio}, {Dencheva}, {Devillepoix}, {Dietrich}, {Eigenbrot}, {Erben}, {Ferreira}, {Foreman-Mackey}, {Fox}, {Freij}, {Garg}, {Geda}, {Glattly}, {Gondhalekar}, {Gordon}, {Grant}, {Greenfield}, {Groener}, {Guest}, {Gurovich}, {Handberg}, {Hart}, {Hatfield-Dodds}, {Homeier}, {Hosseinzadeh}, {Jenness}, {Jones}, {Joseph}, {Kalmbach}, {Karamehmetoglu}, {Ka{\l}uszy{\'n}ski}, {Kelley}, {Kern}, {Kerzendorf}, {Koch}, {Kulumani}, {Lee}, {Ly}, {Ma}, {MacBride}, {Maljaars}, {Muna}, {Murphy}, {Norman},
  {O'Steen}, {Oman}, {Pacifici}, {Pascual}, {Pascual-Granado}, {Patil}, {Perren}, {Pickering}, {Rastogi}, {Roulston}, {Ryan}, {Rykoff}, {Sabater}, {Sakurikar}, {Salgado}, {Sanghi}, {Saunders}, {Savchenko}, {Schwardt}, {Seifert-Eckert}, {Shih}, {Jain}, {Shukla}, {Sick}, {Simpson}, {Singanamalla}, {Singer}, {Singhal}, {Sinha}, {Sip{\H{o}}cz}, {Spitler}, {Stansby}, {Streicher}, {{\v{S}}umak}, {Swinbank}, {Taranu}, {Tewary}, {Tremblay}, {de Val-Borro}, {Van Kooten}, {Vasovi{\'c}}, {Verma}, {de Miranda Cardoso}, {Williams}, {Wilson}, {Winkel}, {Wood-Vasey}, {Xue}, {Yoachim}, {Zhang}, {Zonca}, \& {Astropy Project Contributors}}]{astropy}
{Astropy Collaboration}, {Price-Whelan}, A.~M., {Lim}, P.~L., {et~al.} 2022, \apj, 935, 167, \dodoi{10.3847/1538-4357/ac7c74}

\bibitem[{{Beckwith} \& {Sargent}(1991)}]{beckwith91}
{Beckwith}, S., \& {Sargent}, A. 1991, \apj, 381, 250, \dodoi{10.1086/170646}

\bibitem[{{Beckwith} {et~al.}(1990){Beckwith}, {Sargent}, {Chini}, \& {Guesten}}]{beckwith90}
{Beckwith}, S., {Sargent}, A.~I., {Chini}, R.~S., \& {Guesten}, R. 1990, \aj, 99, 924, \dodoi{10.1086/115385}

\bibitem[{{Beichman} {et~al.}(1988){Beichman}, {Neugebauer}, {Habing}, {Clegg}, \& {Chester}}]{beichman88}
{Beichman}, C., {Neugebauer}, G., {Habing}, H.~J., {Clegg}, P.~E., \& {Chester}, T.~J., eds. 1988, {Infrared astronomical satellite (IRAS) catalogs and atlases. Volume 1: Explanatory supplement}, Vol.~1

\bibitem[{{Birnstiel} {et~al.}(2018){Birnstiel}, {Dullemond}, {Zhu}, {Andrews}, {Bai}, {Wilner}, {Carpenter}, {Huang}, {Isella}, {Benisty}, {P{\'e}rez}, \& {Zhang}}]{dsharp5}
{Birnstiel}, T., {Dullemond}, C.~P., {Zhu}, Z., {et~al.} 2018, \apjl, 869, L45, \dodoi{10.3847/2041-8213/aaf743}

\bibitem[{{Bruggeman}(1935)}]{bruggeman35}
{Bruggeman}, D.~A.~G. 1935, Annalen der Physik, 416, 636, \dodoi{10.1002/andp.19354160705}

\bibitem[{{Carrasco-Gonz{\'a}lez} {et~al.}(2019){Carrasco-Gonz{\'a}lez}, {Sierra}, {Flock}, {Zhu}, {Henning}, {Chandler}, {Galv{\'a}n-Madrid}, {Mac{\'\i}as}, {Anglada}, {Linz}, {Osorio}, {Rodr{\'\i}guez}, {Testi}, {Torrelles}, {P{\'e}rez}, \& {Liu}}]{carrasco-gonzalez19}
{Carrasco-Gonz{\'a}lez}, C., {Sierra}, A., {Flock}, M., {et~al.} 2019, \apj, 883, 71, \dodoi{10.3847/1538-4357/ab3d33}

\bibitem[{{CASA~Team} {et~al.}(2022){CASA~Team}, {Bean}, {Bhatnagar}, {Castro}, {Donovan Meyer}, {Emonts}, {Garcia}, {Garwood}, {Golap}, {Gonzalez Villalba}, {Harris}, {Hayashi}, {Hoskins}, {Hsieh}, {Jagannathan}, {Kawasaki}, {Keimpema}, {Kettenis}, {Lopez}, {Marvil}, {Masters}, {McNichols}, {Mehringer}, {Miel}, {Moellenbrock}, {Montesino}, {Nakazato}, {Ott}, {Petry}, {Pokorny}, {Raba}, {Rau}, {Schiebel}, {Schweighart}, {Sekhar}, {Shimada}, {Small}, {Steeb}, {Sugimoto}, {Suoranta}, {Tsutsumi}, {van Bemmel}, {Verkouter}, {Wells}, {Xiong}, {Szomoru}, {Griffith}, {Glendenning}, \& {Kern}}]{CASA22}
{CASA~Team}, {Bean}, B., {Bhatnagar}, S., {et~al.} 2022, \pasp, 134, 114501, \dodoi{10.1088/1538-3873/ac9642}

\bibitem[{{Chiang} {et~al.}(1996){Chiang}, {Phillips}, \& {Lonsdale}}]{chiang96}
{Chiang}, E., {Phillips}, R.~B., \& {Lonsdale}, C.~J. 1996, \aj, 111, 355, \dodoi{10.1086/117788}

\bibitem[{{Chung} {et~al.}(2024){Chung}, {Andrews}, {Gurwell}, {Wright}, {Long}, {Xu}, \& {Liu}}]{chung24}
{Chung}, C.-Y., {Andrews}, S.~M., {Gurwell}, M.~A., {et~al.} 2024, \apjs, 273, 29, \dodoi{10.3847/1538-4365/ad528b}

\bibitem[{{Chung} {et~al.}(2025){Chung}, {Tsai}, {Wright}, {Xu}, {Long}, {Gurwell}, \& {Liu}}]{chung25}
{Chung}, C.-Y., {Tsai}, A.-L., {Wright}, M., {et~al.} 2025, \apjs, 277, 45, \dodoi{10.3847/1538-4365/adb717}

\bibitem[{{Clarke} {et~al.}(2018){Clarke}, {Tazzari}, {Juhasz}, {Rosotti}, {Booth}, {Facchini}, {Ilee}, {Johns-Krull}, {Kama}, {Meru}, \& {Prato}}]{clarke18}
{Clarke}, C., {Tazzari}, M., {Juhasz}, A., {et~al.} 2018, \apjl, 866, L6, \dodoi{10.3847/2041-8213/aae36b}

\bibitem[{{Coughlan} {et~al.}(2017){Coughlan}, {Ainsworth}, {Eisl{\"o}ffel}, {Hoeft}, {Drabent}, {Scaife}, {Ray}, {Bell}, {Broderick}, {Corbel}, {Grie{\ss}meier}, {van der Horst}, {van Leeuwen}, {Markoff}, {Pietka}, {Stewart}, {Wijers}, \& {Zarka}}]{coughlan17}
{Coughlan}, C.~P., {Ainsworth}, R.~E., {Eisl{\"o}ffel}, J., {et~al.} 2017, \apj, 834, 206, \dodoi{10.3847/1538-4357/834/2/206}

\bibitem[{{D'Alessio} {et~al.}(2006){D'Alessio}, {Calvet}, {Hartmann}, {Franco-Hern{\'a}ndez}, \& {Serv{\'{\i}}n}}]{dalessio06}
{D'Alessio}, P., {Calvet}, N., {Hartmann}, L., {Franco-Hern{\'a}ndez}, R., \& {Serv{\'{\i}}n}, H. 2006, \apj, 638, 314, \dodoi{10.1086/498861}

\bibitem[{Dominik {et~al.}(2021)Dominik, Min, \& Tazaki}]{dominik2021optool}
Dominik, C., Min, M., \& Tazaki, R. 2021, Astrophysics Source Code Library, ascl

\bibitem[{{Draine}(2006)}]{draine06}
{Draine}, B.~T. 2006, \apj, 636, 1114, \dodoi{10.1086/498130}

\bibitem[{{Dr{\k{a}}{\.z}kowska} {et~al.}(2023){Dr{\k{a}}{\.z}kowska}, {Bitsch}, {Lambrechts}, {Mulders}, {Harsono}, {Vazan}, {Liu}, {Ormel}, {Kretke}, \& {Morbidelli}}]{drazkowska23}
{Dr{\k{a}}{\.z}kowska}, J., {Bitsch}, B., {Lambrechts}, M., {et~al.} 2023, in Astronomical Society of the Pacific Conference Series, Vol. 534, Protostars and Planets VII, ed. S.~{Inutsuka}, Y.~{Aikawa}, T.~{Muto}, K.~{Tomida}, \& M.~{Tamura}, 717, \dodoi{10.48550/arXiv.2203.09759}

\bibitem[{{Dutrey} {et~al.}(1996){Dutrey}, {Guilloteau}, {Duvert}, {Prato}, {Simon}, {Schuster}, \& {Menard}}]{dutrey96}
{Dutrey}, A., {Guilloteau}, S., {Duvert}, G., {et~al.} 1996, \aap, 309, 493

\bibitem[{{Duvert} {et~al.}(2000){Duvert}, {Guilloteau}, {M{\'e}nard}, {Simon}, \& {Dutrey}}]{duvert00}
{Duvert}, G., {Guilloteau}, S., {M{\'e}nard}, F., {Simon}, M., \& {Dutrey}, A. 2000, \aap, 355, 165

\bibitem[{{Dzib} {et~al.}(2015){Dzib}, {Loinard}, {Rodr{\'\i}guez}, {Mioduszewski}, {Ortiz-Le{\'o}n}, {Kounkel}, {Pech}, {Rivera}, {Torres}, {Boden}, {Hartmann}, {Evans}, {Brice{\~n}o}, \& {Tobin}}]{dzib15}
{Dzib}, S., {Loinard}, L., {Rodr{\'\i}guez}, L.~F., {et~al.} 2015, \apj, 801, 91, \dodoi{10.1088/0004-637X/801/2/91}

\bibitem[{{Espaillat} {et~al.}(2019){Espaillat}, {Mac{\'\i}as}, {Hern{\'a}ndez}, \& {Robinson}}]{espaillat19}
{Espaillat}, C., {Mac{\'\i}as}, E., {Hern{\'a}ndez}, J., \& {Robinson}, C. 2019, \apjl, 877, L34, \dodoi{10.3847/2041-8213/ab2193}

\bibitem[{Foreman-Mackey(2016)}]{corner}
Foreman-Mackey, D. 2016, The Journal of Open Source Software, 24, \dodoi{10.21105/joss.00024}

\bibitem[{{Foreman-Mackey} {et~al.}(2013){Foreman-Mackey}, {Hogg}, {Lang}, \& {Goodman}}]{foreman-mackey13}
{Foreman-Mackey}, D., {Hogg}, D.~W., {Lang}, D., \& {Goodman}, J. 2013, \pasp, 125, 306, \dodoi{10.1086/670067}

\bibitem[{{Gaia Collaboration} {et~al.}(2023){Gaia Collaboration}, {Vallenari}, {Brown}, {Prusti}, {de Bruijne}, {Arenou}, {Babusiaux}, {Biermann}, {Creevey}, {Ducourant}, {Evans}, {Eyer}, {Guerra}, {Hutton}, {Jordi}, {Klioner}, {Lammers}, {Lindegren}, {Luri}, {Mignard}, {Panem}, {Pourbaix}, {Randich}, {Sartoretti}, {Soubiran}, {Tanga}, {Walton}, {Bailer-Jones}, {Bastian}, {Drimmel}, {Jansen}, {Katz}, {Lattanzi}, {van Leeuwen}, {Bakker}, {Cacciari}, {Casta{\~n}eda}, {De Angeli}, {Fabricius}, {Fouesneau}, {Fr{\'e}mat}, {Galluccio}, {Guerrier}, {Heiter}, {Masana}, {Messineo}, {Mowlavi}, {Nicolas}, {Nienartowicz}, {Pailler}, {Panuzzo}, {Riclet}, {Roux}, {Seabroke}, {Sordo}, {Th{\'e}venin}, {Gracia-Abril}, {Portell}, {Teyssier}, {Altmann}, {Andrae}, {Audard}, {Bellas-Velidis}, {Benson}, {Berthier}, {Blomme}, {Burgess}, {Busonero}, {Busso}, {C{\'a}novas}, {Carry}, {Cellino}, {Cheek}, {Clementini}, {Damerdji}, {Davidson}, {de Teodoro}, {Nu{\~n}ez Campos}, {Delchambre}, {Dell'Oro}, {Esquej},
  {Fern{\'a}ndez-Hern{\'a}ndez}, {Fraile}, {Garabato}, {Garc{\'\i}a-Lario}, {Gosset}, {Haigron}, {Halbwachs}, {Hambly}, {Harrison}, {Hern{\'a}ndez}, {Hestroffer}, {Hodgkin}, {Holl}, {Jan{\ss}en}, {Jevardat de Fombelle}, {Jordan}, {Krone-Martins}, {Lanzafame}, {L{\"o}ffler}, {Marchal}, {Marrese}, {Moitinho}, {Muinonen}, {Osborne}, {Pancino}, {Pauwels}, {Recio-Blanco}, {Reyl{\'e}}, {Riello}, {Rimoldini}, {Roegiers}, {Rybizki}, {Sarro}, {Siopis}, {Smith}, {Sozzetti}, {Utrilla}, {van Leeuwen}, {Abbas}, {{\'A}brah{\'a}m}, {Abreu Aramburu}, {Aerts}, {Aguado}, {Ajaj}, {Aldea-Montero}, {Altavilla}, {{\'A}lvarez}, {Alves}, {Anders}, {Anderson}, {Anglada Varela}, {Antoja}, {Baines}, {Baker}, {Balaguer-N{\'u}{\~n}ez}, {Balbinot}, {Balog}, {Barache}, {Barbato}, {Barros}, {Barstow}, {Bartolom{\'e}}, {Bassilana}, {Bauchet}, {Becciani}, {Bellazzini}, {Berihuete}, {Bernet}, {Bertone}, {Bianchi}, {Binnenfeld}, {Blanco-Cuaresma}, {Blazere}, {Boch}, {Bombrun}, {Bossini}, {Bouquillon}, {Bragaglia}, {Bramante}, {Breedt},
  {Bressan}, {Brouillet}, {Brugaletta}, {Bucciarelli}, {Burlacu}, {Butkevich}, {Buzzi}, {Caffau}, {Cancelliere}, {Cantat-Gaudin}, {Carballo}, {Carlucci}, {Carnerero}, {Carrasco}, {Casamiquela}, {Castellani}, {Castro-Ginard}, {Chaoul}, {Charlot}, {Chemin}, {Chiaramida}, {Chiavassa}, {Chornay}, {Comoretto}, {Contursi}, {Cooper}, {Cornez}, {Cowell}, {Crifo}, {Cropper}, {Crosta}, {Crowley}, {Dafonte}, {Dapergolas}, {David}, {David}, {de Laverny}, {De Luise}, {De March}, {De Ridder}, {de Souza}, {de Torres}, {del Peloso}, {del Pozo}, {Delbo}, {Delgado}, {Delisle}, {Demouchy}, {Dharmawardena}, {Di Matteo}, {Diakite}, {Diener}, {Distefano}, {Dolding}, {Edvardsson}, {Enke}, {Fabre}, {Fabrizio}, {Faigler}, {Fedorets}, {Fernique}, {Fienga}, {Figueras}, {Fournier}, {Fouron}, {Fragkoudi}, {Gai}, {Garcia-Gutierrez}, {Garcia-Reinaldos}, {Garc{\'\i}a-Torres}, {Garofalo}, {Gavel}, {Gavras}, {Gerlach}, {Geyer}, {Giacobbe}, {Gilmore}, {Girona}, {Giuffrida}, {Gomel}, {Gomez}, {Gonz{\'a}lez-N{\'u}{\~n}ez},
  {Gonz{\'a}lez-Santamar{\'\i}a}, {Gonz{\'a}lez-Vidal}, {Granvik}, {Guillout}, {Guiraud}, {Guti{\'e}rrez-S{\'a}nchez}, {Guy}, {Hatzidimitriou}, {Hauser}, {Haywood}, {Helmer}, {Helmi}, {Sarmiento}, {Hidalgo}, {Hilger}, {H{\l}adczuk}, {Hobbs}, {Holland}, {Huckle}, {Jardine}, {Jasniewicz}, {Jean-Antoine Piccolo}, {Jim{\'e}nez-Arranz}, {Jorissen}, {Juaristi Campillo}, {Julbe}, {Karbevska}, {Kervella}, {Khanna}, {Kontizas}, {Kordopatis}, {Korn}, {K{\'o}sp{\'a}l}, {Kostrzewa-Rutkowska}, {Kruszy{\'n}ska}, {Kun}, {Laizeau}, {Lambert}, {Lanza}, {Lasne}, {Le Campion}, {Lebreton}, {Lebzelter}, {Leccia}, {Leclerc}, {Lecoeur-Taibi}, {Liao}, {Licata}, {Lindstr{\o}m}, {Lister}, {Livanou}, {Lobel}, {Lorca}, {Loup}, {Madrero Pardo}, {Magdaleno Romeo}, {Managau}, {Mann}, {Manteiga}, {Marchant}, {Marconi}, {Marcos}, {Marcos Santos}, {Mar{\'\i}n Pina}, {Marinoni}, {Marocco}, {Marshall}, {Martin Polo}, {Mart{\'\i}n-Fleitas}, {Marton}, {Mary}, {Masip}, {Massari}, {Mastrobuono-Battisti}, {Mazeh}, {McMillan}, {Messina}, {Michalik},
  {Millar}, {Mints}, {Molina}, {Molinaro}, {Moln{\'a}r}, {Monari}, {Mongui{\'o}}, {Montegriffo}, {Montero}, {Mor}, {Mora}, {Morbidelli}, {Morel}, {Morris}, {Muraveva}, {Murphy}, {Musella}, {Nagy}, {Noval}, {Oca{\~n}a}, {Ogden}, {Ordenovic}, {Osinde}, {Pagani}, {Pagano}, {Palaversa}, {Palicio}, {Pallas-Quintela}, {Panahi}, {Payne-Wardenaar}, {Pe{\~n}alosa Esteller}, {Penttil{\"a}}, {Pichon}, {Piersimoni}, {Pineau}, {Plachy}, {Plum}, {Poggio}, {Pr{\v{s}}a}, {Pulone}, {Racero}, {Ragaini}, {Rainer}, {Raiteri}, {Rambaux}, {Ramos}, {Ramos-Lerate}, {Re Fiorentin}, {Regibo}, {Richards}, {Rios Diaz}, {Ripepi}, {Riva}, {Rix}, {Rixon}, {Robichon}, {Robin}, {Robin}, {Roelens}, {Rogues}, {Rohrbasser}, {Romero-G{\'o}mez}, {Rowell}, {Royer}, {Ruz Mieres}, {Rybicki}, {Sadowski}, {S{\'a}ez N{\'u}{\~n}ez}, {Sagrist{\`a} Sell{\'e}s}, {Sahlmann}, {Salguero}, {Samaras}, {Sanchez Gimenez}, {Sanna}, {Santove{\~n}a}, {Sarasso}, {Schultheis}, {Sciacca}, {Segol}, {Segovia}, {S{\'e}gransan}, {Semeux}, {Shahaf}, {Siddiqui}, {Siebert},
  {Siltala}, {Silvelo}, {Slezak}, {Slezak}, {Smart}, {Snaith}, {Solano}, {Solitro}, {Souami}, {Souchay}, {Spagna}, {Spina}, {Spoto}, {Steele}, {Steidelm{\"u}ller}, {Stephenson}, {S{\"u}veges}, {Surdej}, {Szabados}, {Szegedi-Elek}, {Taris}, {Taylor}, {Teixeira}, {Tolomei}, {Tonello}, {Torra}, {Torra}, {Torralba Elipe}, {Trabucchi}, {Tsounis}, {Turon}, {Ulla}, {Unger}, {Vaillant}, {van Dillen}, {van Reeven}, {Vanel}, {Vecchiato}, {Viala}, {Vicente}, {Voutsinas}, {Weiler}, {Wevers}, {Wyrzykowski}, {Yoldas}, {Yvard}, {Zhao}, {Zorec}, {Zucker}, \& {Zwitter}}]{gaia_dr3}
{Gaia Collaboration}, {Vallenari}, A., {Brown}, A.~G.~A., {et~al.} 2023, \aap, 674, A1, \dodoi{10.1051/0004-6361/202243940}

\bibitem[{{Gangi} {et~al.}(2022){Gangi}, {Antoniucci}, {Biazzo}, {Frasca}, {Nisini}, {Alcal{\'a}}, {Giannini}, {Manara}, {Giunta}, {Harutyunyan}, {Munari}, \& {Vitali}}]{gangi22}
{Gangi}, M., {Antoniucci}, S., {Biazzo}, K., {et~al.} 2022, \aap, 667, A124, \dodoi{10.1051/0004-6361/202244042}

\bibitem[{{Garufi} {et~al.}(2025){Garufi}, {Carrasco-Gonz{\'a}lez}, {Mac{\'\i}as}, {Testi}, {Curone}, {Ricci}, {Facchini}, {Long}, {Manara}, {Pascucci}, {Rosotti}, {Zagaria}, {Clarke}, {Herczeg}, {Isella}, {Rota}, {Mauc{\'o}}, {van der Marel}, \& {Tazzari}}]{garufi25}
{Garufi}, A., {Carrasco-Gonz{\'a}lez}, C., {Mac{\'\i}as}, E., {et~al.} 2025, \aap, 694, A290, \dodoi{10.1051/0004-6361/202452496}

\bibitem[{{Godines} {et~al.}(2025){Godines}, {Lyra}, {Ricci}, {Yang}, {Simon}, {Lim}, \& {Carrera}}]{godines25}
{Godines}, D., {Lyra}, W., {Ricci}, L., {et~al.} 2025, arXiv e-prints, arXiv:2506.10435, \dodoi{10.48550/arXiv.2506.10435}

\bibitem[{{Goodman} \& {Weare}(2010)}]{goodman10}
{Goodman}, J., \& {Weare}, J. 2010, Comm.~App.~Math.~Comp.~Sci., 5, 65

\bibitem[{{Greaves} \& {Mason}(2022)}]{greaves22}
{Greaves}, J., \& {Mason}, B. 2022, \mnras, 513, 3180, \dodoi{10.1093/mnras/stac856}

\bibitem[{{Greaves} \& {Rice}(2010)}]{greaves10}
{Greaves}, J.~S., \& {Rice}, W.~K.~M. 2010, \mnras, 407, 1981, \dodoi{10.1111/j.1365-2966.2010.17043.x}

\bibitem[{{G{\"u}del}(2002)}]{guedel02}
{G{\"u}del}, M. 2002, \araa, 40, 217, \dodoi{10.1146/annurev.astro.40.060401.093806}

\bibitem[{Guidi {et~al.}(2022)Guidi, Isella, Testi, Chandler, Liu, Schmid, Rosotti, Meng, Jennings, Williams, {et~al.}}]{guidi22}
Guidi, G., Isella, A., Testi, L., {et~al.} 2022, Astronomy \& Astrophysics, 664, A137

\bibitem[{{Guilloteau} {et~al.}(2011){Guilloteau}, {Dutrey}, {Pi{\'e}tu}, \& {Boehler}}]{guilloteau11}
{Guilloteau}, S., {Dutrey}, A., {Pi{\'e}tu}, V., \& {Boehler}, Y. 2011, \aap, 529, A105, \dodoi{10.1051/0004-6361/201015209}

\bibitem[{{Hamidouche} {et~al.}(2006){Hamidouche}, {Looney}, \& {Mundy}}]{hamidouche06}
{Hamidouche}, M., {Looney}, L.~W., \& {Mundy}, L.~G. 2006, \apj, 651, 321, \dodoi{10.1086/507693}

\bibitem[{Harris {et~al.}(2020)Harris, Millman, van~der Walt, Gommers, Virtanen, Cournapeau, Wieser, Taylor, Berg, Smith, Kern, Picus, Hoyer, van Kerkwijk, Brett, Haldane, del R{\'{i}}o, Wiebe, Peterson, G{\'{e}}rard-Marchant, Sheppard, Reddy, Weckesser, Abbasi, Gohlke, \& Oliphant}]{numpy}
Harris, C.~R., Millman, K.~J., van~der Walt, S.~J., {et~al.} 2020, Nature, 585, 357, \dodoi{10.1038/s41586-020-2649-2}

\bibitem[{{Harrison} {et~al.}(2024){Harrison}, {Lin}, {Looney}, {Li}, {Yang}, {Stephens}, \& {Fern{\'a}ndez-L{\'o}pez}}]{harrison24}
{Harrison}, R., {Lin}, Z.-Y.~D., {Looney}, L.~W., {et~al.} 2024, \apj, 967, 40, \dodoi{10.3847/1538-4357/ad39ec}

\bibitem[{{Harrison} {et~al.}(2019){Harrison}, {Looney}, {Stephens}, {Li}, {Yang}, {Kataoka}, {Harris}, {Kwon}, {Muto}, \& {Momose}}]{harrison19}
{Harrison}, R., {Looney}, L.~W., {Stephens}, I.~W., {et~al.} 2019, \apjl, 877, L2, \dodoi{10.3847/2041-8213/ab1e46}

\bibitem[{{Hartigan} {et~al.}(1995){Hartigan}, {Edwards}, \& {Ghandour}}]{hartigan95}
{Hartigan}, P., {Edwards}, S., \& {Ghandour}, L. 1995, \apj, 452, 736, \dodoi{10.1086/176344}

\bibitem[{{Hashimoto} {et~al.}(2022){Hashimoto}, {Liu}, {Dong}, {Liu}, \& {Muto}}]{hashimoto22}
{Hashimoto}, J., {Liu}, H.~B., {Dong}, R., {Liu}, B., \& {Muto}, T. 2022, \apj, 941, 66, \dodoi{10.3847/1538-4357/aca01d}

\bibitem[{{Henning} \& {Stognienko}(1996)}]{henning96}
{Henning}, T., \& {Stognienko}, R. 1996, \aap, 311, 291

\bibitem[{{Hildebrand}(1983)}]{hildebrand83}
{Hildebrand}, R.~H. 1983, \qjras, 24, 267

\bibitem[{{Ho} {et~al.}(2004){Ho}, {Moran}, \& {Lo}}]{ho04}
{Ho}, P., {Moran}, J.~M., \& {Lo}, K.~Y. 2004, \apjl, 616, L1, \dodoi{10.1086/423245}

\bibitem[{{Hoang} {et~al.}(2018){Hoang}, {Lan}, {Vinh}, \& {Kim}}]{hoang18}
{Hoang}, T., {Lan}, N.-Q., {Vinh}, N.-A., \& {Kim}, Y.-J. 2018, \apj, 862, 116, \dodoi{10.3847/1538-4357/aaccf0}

\bibitem[{{Howard} {et~al.}(2013){Howard}, {Sandell}, {Vacca}, {Duch{\^e}ne}, {Mathews}, {Augereau}, {Barrado}, {Dent}, {Eiroa}, {Grady}, {Kamp}, {Meeus}, {M{\'e}nard}, {Pinte}, {Podio}, {Riviere-Marichalar}, {Roberge}, {Thi}, {Vicente}, \& {Williams}}]{howard13}
{Howard}, C., {Sandell}, G., {Vacca}, W.~D., {et~al.} 2013, \apj, 776, 21, \dodoi{10.1088/0004-637X/776/1/21}

\bibitem[{{Huang} {et~al.}(2023){Huang}, {Bergin}, {Bae}, {Benisty}, \& {Andrews}}]{huang23}
{Huang}, J., {Bergin}, E.~A., {Bae}, J., {Benisty}, M., \& {Andrews}, S.~M. 2023, \apj, 943, 107, \dodoi{10.3847/1538-4357/aca89c}

\bibitem[{{Huang} {et~al.}(2017){Huang}, {{\"O}berg}, {Qi}, {Aikawa}, {Andrews}, {Furuya}, {Guzm{\'a}n}, {Loomis}, {van Dishoeck}, \& {Wilner}}]{huang17}
{Huang}, J., {{\"O}berg}, K.~I., {Qi}, C., {et~al.} 2017, \apj, 835, 231, \dodoi{10.3847/1538-4357/835/2/231}

\bibitem[{{Huang} {et~al.}(2018{\natexlab{a}}){Huang}, {Andrews}, {Dullemond}, {Isella}, {P{\'e}rez}, {Guzm{\'a}n}, {{\"O}berg}, {Zhu}, {Zhang}, {Bai}, {Benisty}, {Birnstiel}, {Carpenter}, {Hughes}, {Ricci}, {Weaver}, \& {Wilner}}]{dsharp2}
{Huang}, J., {Andrews}, S.~M., {Dullemond}, C.~P., {et~al.} 2018{\natexlab{a}}, \apjl, 869, L42, \dodoi{10.3847/2041-8213/aaf740}

\bibitem[{{Huang} {et~al.}(2018{\natexlab{b}}){Huang}, {Andrews}, {Cleeves}, {{\"O}berg}, {Wilner}, {Bai}, {Birnstiel}, {Carpenter}, {Hughes}, {Isella}, {P{\'e}rez}, {Ricci}, \& {Zhu}}]{huang18}
{Huang}, J., {Andrews}, S.~M., {Cleeves}, L.~I., {et~al.} 2018{\natexlab{b}}, \apj, 852, 122, \dodoi{10.3847/1538-4357/aaa1e7}

\bibitem[{{Huang} {et~al.}(2020){Huang}, {Andrews}, {Dullemond}, {{\"O}berg}, {Qi}, {Zhu}, {Birnstiel}, {Carpenter}, {Isella}, {Mac{\'\i}as}, {McClure}, {P{\'e}rez}, {Teague}, {Wilner}, \& {Zhang}}]{huang20}
{Huang}, J., {Andrews}, S.~M., {Dullemond}, C.~P., {et~al.} 2020, \apj, 891, 48, \dodoi{10.3847/1538-4357/ab711e}

\bibitem[{{Hughes} {et~al.}(2008){Hughes}, {Wilner}, {Qi}, \& {Hogerheijde}}]{hughes08}
{Hughes}, A., {Wilner}, D.~J., {Qi}, C., \& {Hogerheijde}, M.~R. 2008, \apj, 678, 1119, \dodoi{10.1086/586730}

\bibitem[{{Hughes} {et~al.}(2009){Hughes}, {Andrews}, {Espaillat}, {Wilner}, {Calvet}, {D'Alessio}, {Qi}, {Williams}, \& {Hogerheijde}}]{hughes09}
{Hughes}, A., {Andrews}, S.~M., {Espaillat}, C., {et~al.} 2009, \apj, 698, 131, \dodoi{10.1088/0004-637X/698/1/131}

\bibitem[{Hunter(2007)}]{matplotlib}
Hunter, J.~D. 2007, Computing In Science \& Engineering, 9, 90

\bibitem[{{Isella} {et~al.}(2009){Isella}, {Carpenter}, \& {Sargent}}]{isella09}
{Isella}, A., {Carpenter}, J.~M., \& {Sargent}, A.~I. 2009, \apj, 701, 260, \dodoi{10.1088/0004-637X/701/1/260}

\bibitem[{{Isella} {et~al.}(2014){Isella}, {Chandler}, {Carpenter}, {P{\'e}rez}, \& {Ricci}}]{isella14}
{Isella}, A., {Chandler}, C.~J., {Carpenter}, J.~M., {P{\'e}rez}, L.~M., \& {Ricci}, L. 2014, \apj, 788, 129, \dodoi{10.1088/0004-637X/788/2/129}

\bibitem[{{Isella} {et~al.}(2012){Isella}, {P{\'e}rez}, \& {Carpenter}}]{isella12}
{Isella}, A., {P{\'e}rez}, L.~M., \& {Carpenter}, J.~M. 2012, \apj, 747, 136, \dodoi{10.1088/0004-637X/747/2/136}

\bibitem[{{Ishihara} {et~al.}(2010){Ishihara}, {Onaka}, {Kataza}, {Salama}, {Alfageme}, {Cassatella}, {Cox}, {Garc{\'{\i}}a-Lario}, {Stephenson}, {Cohen}, {Fujishiro}, {Fujiwara}, {Hasegawa}, {Ita}, {Kim}, {Matsuhara}, {Murakami}, {M{\"u}ller}, {Nakagawa}, {Ohyama}, {Oyabu}, {Pyo}, {Sakon}, {Shibai}, {Takita}, {Tanab{\'e}}, {Uemizu}, {Ueno}, {Usui}, {Wada}, {Watarai}, {Yamamura}, \& {Yamauchi}}]{ishihari10}
{Ishihara}, D., {Onaka}, T., {Kataza}, H., {et~al.} 2010, \aap, 514, A1, \dodoi{10.1051/0004-6361/200913811}

\bibitem[{{Jennings} {et~al.}(2022){Jennings}, {Tazzari}, {Clarke}, {Booth}, \& {Rosotti}}]{jennings22}
{Jennings}, J., {Tazzari}, M., {Clarke}, C.~J., {Booth}, R.~A., \& {Rosotti}, G.~P. 2022, \mnras, 514, 6053, \dodoi{10.1093/mnras/stac1770}

\bibitem[{{Kataoka} {et~al.}(2014){Kataoka}, {Okuzumi}, {Tanaka}, \& {Nomura}}]{kataoka14}
{Kataoka}, A., {Okuzumi}, S., {Tanaka}, H., \& {Nomura}, H. 2014, \aap, 568, A42, \dodoi{10.1051/0004-6361/201323199}

\bibitem[{{Kitamura} {et~al.}(2002){Kitamura}, {Momose}, {Yokogawa}, {Kawabe}, {Tamura}, \& {Ida}}]{kitamura02}
{Kitamura}, Y., {Momose}, M., {Yokogawa}, S., {et~al.} 2002, \apj, 581, 357, \dodoi{10.1086/344223}

\bibitem[{{Kwon} {et~al.}(2015){Kwon}, {Looney}, {Mundy}, \& {Welch}}]{kwon15}
{Kwon}, W., {Looney}, L.~W., {Mundy}, L.~G., \& {Welch}, W.~J. 2015, \apj, 808, 102, \dodoi{10.1088/0004-637X/808/1/102}

\bibitem[{{Liu}(2019)}]{liu19}
{Liu}, H.~B. 2019, \apjl, 877, L22, \dodoi{10.3847/2041-8213/ab1f8e}

\bibitem[{{Liu} {et~al.}(2024{\natexlab{a}}){Liu}, {Casassus}, {Dong}, {Doi}, {Hashimoto}, \& {Muto}}]{liu24}
{Liu}, H.~B., {Casassus}, S., {Dong}, R., {et~al.} 2024{\natexlab{a}}, \apj, 972, 163, \dodoi{10.3847/1538-4357/ad5dab}

\bibitem[{{Liu} {et~al.}(2024{\natexlab{b}}){Liu}, {Muto}, {Konishi}, {Chung}, {Hashimoto}, {Doi}, {Dong}, {Kudo}, {Hasegawa}, {Terada}, \& {Kataoka}}]{liu24_dmtau}
{Liu}, H.~B., {Muto}, T., {Konishi}, M., {et~al.} 2024{\natexlab{b}}, \aap, 685, A18, \dodoi{10.1051/0004-6361/202348896}

\bibitem[{{Long} {et~al.}(2018){Long}, {Pinilla}, {Herczeg}, {Harsono}, {Dipierro}, {Pascucci}, {Hendler}, {Tazzari}, {Ragusa}, {Salyk}, {Edwards}, {Lodato}, {van de Plas}, {Johnstone}, {Liu}, {Boehler}, {Cabrit}, {Manara}, {Menard}, {Mulders}, {Nisini}, {Fischer}, {Rigliaco}, {Banzatti}, {Avenhaus}, \& {Gully-Santiago}}]{long18}
{Long}, F., {Pinilla}, P., {Herczeg}, G.~J., {et~al.} 2018, \apj, 869, 17, \dodoi{10.3847/1538-4357/aae8e1}

\bibitem[{{Long} {et~al.}(2019){Long}, {Herczeg}, {Harsono}, {Pinilla}, {Tazzari}, {Manara}, {Pascucci}, {Cabrit}, {Nisini}, {Johnstone}, {Edwards}, {Salyk}, {Menard}, {Lodato}, {Boehler}, {Mace}, {Liu}, {Mulders}, {Hendler}, {Ragusa}, {Fischer}, {Banzatti}, {Rigliaco}, {van de Plas}, {Dipierro}, {Gully-Santiago}, \& {Lopez-Valdivia}}]{long19}
{Long}, F., {Herczeg}, G.~J., {Harsono}, D., {et~al.} 2019, \apj, 882, 49, \dodoi{10.3847/1538-4357/ab2d2d}

\bibitem[{{Long} {et~al.}(2020){Long}, {Pinilla}, {Herczeg}, {Andrews}, {Harsono}, {Johnstone}, {Ragusa}, {Pascucci}, {Wilner}, {Hendler}, {Jennings}, {Liu}, {Lodato}, {Menard}, {van de Plas}, \& {Dipierro}}]{long20}
{Long}, F., {Pinilla}, P., {Herczeg}, G.~J., {et~al.} 2020, \apj, 898, 36, \dodoi{10.3847/1538-4357/ab9a54}

\bibitem[{{Looney} {et~al.}(2000){Looney}, {Mundy}, \& {Welch}}]{looney00}
{Looney}, L., {Mundy}, L.~G., \& {Welch}, W.~J. 2000, \apj, 529, 477, \dodoi{10.1086/308239}

\bibitem[{{Mac{\'\i}as} {et~al.}(2021){Mac{\'\i}as}, {Guerra-Alvarado}, {Carrasco-Gonz{\'a}lez}, {Ribas}, {Espaillat}, {Huang}, \& {Andrews}}]{macias21}
{Mac{\'\i}as}, E., {Guerra-Alvarado}, O., {Carrasco-Gonz{\'a}lez}, C., {et~al.} 2021, \aap, 648, A33, \dodoi{10.1051/0004-6361/202039812}

\bibitem[{{Mac{\'\i}as} {et~al.}(2016){Mac{\'\i}as}, {Anglada}, {Osorio}, {Calvet}, {Torrelles}, {G{\'o}mez}, {Espaillat}, {Lizano}, {Rodr{\'\i}guez}, {Carrasco-Gonz{\'a}lez}, \& {Zapata}}]{macias16}
{Mac{\'\i}as}, E., {Anglada}, G., {Osorio}, M., {et~al.} 2016, \apj, 829, 1, \dodoi{10.3847/0004-637X/829/1/1}

\bibitem[{{Manara} {et~al.}(2023){Manara}, {Ansdell}, {Rosotti}, {Hughes}, {Armitage}, {Lodato}, \& {Williams}}]{manara23}
{Manara}, C.~F., {Ansdell}, M., {Rosotti}, G.~P., {et~al.} 2023, in Astronomical Society of the Pacific Conference Series, Vol. 534, Protostars and Planets VII, ed. S.~{Inutsuka}, Y.~{Aikawa}, T.~{Muto}, K.~{Tomida}, \& M.~{Tamura}, 539, \dodoi{10.48550/arXiv.2203.09930}

\bibitem[{{Manara} {et~al.}(2018){Manara}, {Morbidelli}, \& {Guillot}}]{manara18}
{Manara}, C.~F., {Morbidelli}, A., \& {Guillot}, T. 2018, \aap, 618, L3, \dodoi{10.1051/0004-6361/201834076}

\bibitem[{{Mannings} \& {Emerson}(1994)}]{mannings94}
{Mannings}, V., \& {Emerson}, J. 1994, \mnras, 267, 361

\bibitem[{{Mannings} \& {Sargent}(1997)}]{mannings97}
{Mannings}, V., \& {Sargent}, A. 1997, \apj, 490, 792, \dodoi{10.1086/304897}

\bibitem[{{Mezger} \& {Henderson}(1967)}]{mezger67}
{Mezger}, P., \& {Henderson}, A. 1967, \apj, 147, 471, \dodoi{10.1086/149030}

\bibitem[{{Min} {et~al.}(2005){Min}, {Hovenier}, \& {de Koter}}]{min05}
{Min}, M., {Hovenier}, J.~W., \& {de Koter}, A. 2005, \aap, 432, 909, \dodoi{10.1051/0004-6361:20041920}

\bibitem[{{Miyake} \& {Nakagawa}(1993)}]{miyake93}
{Miyake}, K., \& {Nakagawa}, Y. 1993, Icarus, 106, 20, \dodoi{10.1006/icar.1993.1156}

\bibitem[{Miyake \& Nakagawa(1993)}]{miyake1993effects}
Miyake, K., \& Nakagawa, Y. 1993, icarus, 106, 20

\bibitem[{{Najita} \& {Kenyon}(2014)}]{najita14}
{Najita}, J.~R., \& {Kenyon}, S.~J. 2014, \mnras, 445, 3315, \dodoi{10.1093/mnras/stu1994}

\bibitem[{{{\"O}berg} {et~al.}(2010){{\"O}berg}, {Qi}, {Fogel}, {Bergin}, {Andrews}, {Espaillat}, {van Kempen}, {Wilner}, \& {Pascucci}}]{oberg10}
{{\"O}berg}, K., {Qi}, C., {Fogel}, J.~K.~J., {et~al.} 2010, \apj, 720, 480, \dodoi{10.1088/0004-637X/720/1/480}

\bibitem[{{Ohashi} {et~al.}(1996){Ohashi}, {Hayashi}, {Kawabe}, \& {Ishiguro}}]{ohashi96}
{Ohashi}, N., {Hayashi}, M., {Kawabe}, R., \& {Ishiguro}, M. 1996, \apj, 466, 317, \dodoi{10.1086/177512}

\bibitem[{{Osterloh} \& {Beckwith}(1995)}]{osterloh95}
{Osterloh}, M., \& {Beckwith}, S. 1995, \apj, 439, 288, \dodoi{10.1086/175172}

\bibitem[{Paine(2024)}]{paine_am}
Paine, S. 2024, The am atmospheric model, 14.0,  Zenodo, \dodoi{10.5281/zenodo.13748391}

\bibitem[{{Pascucci} {et~al.}(2012){Pascucci}, {Gorti}, \& {Hollenbach}}]{pascucci12}
{Pascucci}, I., {Gorti}, U., \& {Hollenbach}, D. 2012, \apjl, 751, L42, \dodoi{10.1088/2041-8205/751/2/L42}

\bibitem[{{Pascucci} {et~al.}(2014){Pascucci}, {Ricci}, {Gorti}, {Hollenbach}, {Hendler}, {Brooks}, \& {Contreras}}]{pascucci14}
{Pascucci}, I., {Ricci}, L., {Gorti}, U., {et~al.} 2014, \apj, 795, 1, \dodoi{10.1088/0004-637X/795/1/1}

\bibitem[{{Perley} \& {Butler}(2017)}]{perley17}
{Perley}, R., \& {Butler}, B. 2017, \apjs, 230, 7, \dodoi{10.3847/1538-4365/aa6df9}

\bibitem[{{Pi{\'e}tu} {et~al.}(2006){Pi{\'e}tu}, {Dutrey}, {Guilloteau}, {Chapillon}, \& {Pety}}]{pietu06}
{Pi{\'e}tu}, V., {Dutrey}, A., {Guilloteau}, S., {Chapillon}, E., \& {Pety}, J. 2006, \aap, 460, L43, \dodoi{10.1051/0004-6361:20065968}

\bibitem[{{Pollack} {et~al.}(1996){Pollack}, {Hubickyj}, {Bodenheimer}, {Lissauer}, {Podolak}, \& {Greenzweig}}]{pollack96}
{Pollack}, J.~B., {Hubickyj}, O., {Bodenheimer}, P., {et~al.} 1996, Icarus, 124, 62, \dodoi{10.1006/icar.1996.0190}

\bibitem[{{Rafikov}(2006)}]{rafikov06}
{Rafikov}, R.~R. 2006, \apj, 646, 288, \dodoi{10.1086/504793}

\bibitem[{{Rau} \& {Cornwell}(2011)}]{rau11}
{Rau}, U., \& {Cornwell}, T. 2011, \aap, 532, A71, \dodoi{10.1051/0004-6361/201117104}

\bibitem[{{Rebull} {et~al.}(2010){Rebull}, {Padgett}, {McCabe}, {Hillenbrand}, {Stapelfeldt}, {Noriega-Crespo}, {Carey}, {Brooke}, {Huard}, {Terebey}, {Audard}, {Monin}, {Fukagawa}, {G{\"u}del}, {Knapp}, {Menard}, {Allen}, {Angione}, {Baldovin-Saavedra}, {Bouvier}, {Briggs}, {Dougados}, {Evans}, {Flagey}, {Guieu}, {Grosso}, {Glauser}, {Harvey}, {Hines}, {Latter}, {Skinner}, {Strom}, {Tromp}, \& {Wolf}}]{rebull10}
{Rebull}, L., {Padgett}, D.~L., {McCabe}, C.-E., {et~al.} 2010, \apjs, 186, 259, \dodoi{10.1088/0067-0049/186/2/259}

\bibitem[{{Reynolds}(1986)}]{reynolds86}
{Reynolds}, S.~P. 1986, \apj, 304, 713, \dodoi{10.1086/164209}

\bibitem[{Ribas {et~al.}(2020)Ribas, Espaillat, Mac{\'\i}as, \& Sarro}]{ribas2020modeling}
Ribas, {\'A}., Espaillat, C.~C., Mac{\'\i}as, E., \& Sarro, L.~M. 2020, Astronomy \& Astrophysics, 642, A171

\bibitem[{{Ribas} {et~al.}(2017){Ribas}, {Espaillat}, {Mac{\'{\i}}as}, {Bouy}, {Andrews}, {Calvet}, {Naylor}, {Riviere-Marichalar}, {van der Wiel}, \& {Wilner}}]{ribas17}
{Ribas}, {\'A}., {Espaillat}, C.~C., {Mac{\'{\i}}as}, E., {et~al.} 2017, \apj, 849, 63, \dodoi{10.3847/1538-4357/aa8e99}

\bibitem[{{Ricci} {et~al.}(2010{\natexlab{a}}){Ricci}, {Testi}, {Natta}, \& {Brooks}}]{ricci10b}
{Ricci}, L., {Testi}, L., {Natta}, A., \& {Brooks}, K.~J. 2010{\natexlab{a}}, \aap, 521, A66, \dodoi{10.1051/0004-6361/201015039}

\bibitem[{{Ricci} {et~al.}(2010{\natexlab{b}}){Ricci}, {Testi}, {Natta}, {Neri}, {Cabrit}, \& {Herczeg}}]{ricci10a}
{Ricci}, L., {Testi}, L., {Natta}, A., {et~al.} 2010{\natexlab{b}}, \aap, 512, A15, \dodoi{10.1051/0004-6361/200913403}

\bibitem[{{Ricci} {et~al.}(2012){Ricci}, {Trotta}, {Testi}, {Natta}, {Isella}, \& {Wilner}}]{ricci12}
{Ricci}, L., {Trotta}, F., {Testi}, L., {et~al.} 2012, \aap, 540, A6, \dodoi{10.1051/0004-6361/201118296}

\bibitem[{{Rodmann} {et~al.}(2006){Rodmann}, {Henning}, {Chandler}, {Mundy}, \& {Wilner}}]{rodmann06}
{Rodmann}, J., {Henning}, T., {Chandler}, C.~J., {Mundy}, L.~G., \& {Wilner}, D.~J. 2006, \aap, 446, 211, \dodoi{10.1051/0004-6361:20054038}

\bibitem[{{Rota} {et~al.}(2025){Rota}, {van der Marel}, {Garufi}, {Carrasco-Gonz{\'a}lez}, {Macias}, {Pascucci}, {Sellek}, {Testi}, {Isella}, \& {Facchini}}]{rota25}
{Rota}, A.~A., {van der Marel}, N., {Garufi}, A., {et~al.} 2025, arXiv e-prints, arXiv:2505.16586, \dodoi{10.48550/arXiv.2505.16586}

\bibitem[{{Sandell} {et~al.}(2011){Sandell}, {Weintraub}, \& {Hamidouche}}]{sandell11}
{Sandell}, G., {Weintraub}, D.~A., \& {Hamidouche}, M. 2011, \apj, 727, 26, \dodoi{10.1088/0004-637X/727/1/26}

\bibitem[{{Sierra} {et~al.}(2021){Sierra}, {P{\'e}rez}, {Zhang}, {Law}, {Guzm{\'a}n}, {Qi}, {Bosman}, {{\"O}berg}, {Andrews}, {Long}, {Teague}, {Booth}, {Walsh}, {Wilner}, {M{\'e}nard}, {Cataldi}, {Czekala}, {Bae}, {Huang}, {Bergner}, {Ilee}, {Benisty}, {Le Gal}, {Loomis}, {Tsukagoshi}, {Liu}, {Yamato}, \& {Aikawa}}]{sierra21}
{Sierra}, A., {P{\'e}rez}, L.~M., {Zhang}, K., {et~al.} 2021, \apjs, 257, 14, \dodoi{10.3847/1538-4365/ac1431}

\bibitem[{{Sturm} {et~al.}(2022){Sturm}, {McClure}, {Harsono}, {Facchini}, {Long}, {Kama}, {Bergin}, \& {van Dishoeck}}]{sturm22}
{Sturm}, J., {McClure}, M.~K., {Harsono}, D., {et~al.} 2022, \aap, 660, A126, \dodoi{10.1051/0004-6361/202141860}

\bibitem[{{Tazzari} {et~al.}(2016){Tazzari}, {Testi}, {Ercolano}, {Natta}, {Isella}, {Chandler}, {P{\'e}rez}, {Andrews}, {Wilner}, {Ricci}, {Henning}, {Linz}, {Kwon}, {Corder}, {Dullemond}, {Carpenter}, {Sargent}, {Mundy}, {Storm}, {Calvet}, {Greaves}, {Lazio}, \& {Deller}}]{tazzari16}
{Tazzari}, M., {Testi}, L., {Ercolano}, B., {et~al.} 2016, \aap, 588, A53, \dodoi{10.1051/0004-6361/201527423}

\bibitem[{{Tazzari} {et~al.}(2021){Tazzari}, {Testi}, {Natta}, {Williams}, {Ansdell}, {Carpenter}, {Facchini}, {Guidi}, {Hogherheijde}, {Manara}, {Miotello}, \& {van der Marel}}]{tazzari21}
{Tazzari}, M., {Testi}, L., {Natta}, A., {et~al.} 2021, \mnras, 506, 5117, \dodoi{10.1093/mnras/stab1912}

\bibitem[{{Teague} {et~al.}(2025){Teague}, {Benisty}, {Facchini}, {Fukagawa}, {Pinte}, {Andrews}, {Bae}, {Barraza-Alfaro}, {Cataldi}, {Cuello}, {Curone}, {Czekala}, {Fasano}, {Flock}, {Galloway-Sprietsma}, {Garg}, {Hall}, {Hammond}, {Hilder}, {Huang}, {Ilee}, {Izquierdo}, {Kanagawa}, {Lesur}, {Lodato}, {Longarini}, {Loomis}, {Masset}, {Menard}, {Orihara}, {Price}, {Rosotti}, {Stadler}, {Testi}, {Yen}, {Wafflard-Fernandez}, {Wilner}, {Winter}, {W{\"o}lfer}, {Yoshida}, \& {Zawadzki}}]{exoALMA_I}
{Teague}, R., {Benisty}, M., {Facchini}, S., {et~al.} 2025, \apjl, 984, L6, \dodoi{10.3847/2041-8213/adc43b}

\bibitem[{{Throop} \& {Bally}(2008)}]{throop08}
{Throop}, H.~B., \& {Bally}, J. 2008, \aj, 135, 2380, \dodoi{10.1088/0004-6256/135/6/2380}

\bibitem[{{Tremaine} {et~al.}(2002){Tremaine}, {Gebhardt}, {Bender}, {Bower}, {Dressler}, {Faber}, {Filippenko}, {Green}, {Grillmair}, {Ho}, {Kormendy}, {Lauer}, {Magorrian}, {Pinkney}, \& {Richstone}}]{tremaine02}
{Tremaine}, S., {Gebhardt}, K., {Bender}, R., {et~al.} 2002, \apj, 574, 740, \dodoi{10.1086/341002}

\bibitem[{{Tripathi} {et~al.}(2017){Tripathi}, {Andrews}, {Birnstiel}, \& {Wilner}}]{tripathi17}
{Tripathi}, A., {Andrews}, S.~M., {Birnstiel}, T., \& {Wilner}, D.~J. 2017, \apj, 845, 44, \dodoi{10.3847/1538-4357/aa7c62}

\bibitem[{{Tsukagoshi} {et~al.}(2016){Tsukagoshi}, {Nomura}, {Muto}, {Kawabe}, {Ishimoto}, {Kanagawa}, {Okuzumi}, {Ida}, {Walsh}, \& {Millar}}]{tsukagoshi16}
{Tsukagoshi}, T., {Nomura}, H., {Muto}, T., {et~al.} 2016, \apjl, 829, L35, \dodoi{10.3847/2041-8205/829/2/L35}

\bibitem[{{Tychoniec} {et~al.}(2020){Tychoniec}, {Manara}, {Rosotti}, {van Dishoeck}, {Cridland}, {Hsieh}, {Murillo}, {Segura-Cox}, {van Terwisga}, \& {Tobin}}]{tychoniec20}
{Tychoniec}, {\L}., {Manara}, C.~F., {Rosotti}, G.~P., {et~al.} 2020, \aap, 640, A19, \dodoi{10.1051/0004-6361/202037851}

\bibitem[{{Ubach} {et~al.}(2012){Ubach}, {Maddison}, {Wright}, {Wilner}, {Lommen}, \& {Koribalski}}]{ubach12}
{Ubach}, C., {Maddison}, S.~T., {Wright}, C.~M., {et~al.} 2012, \mnras, 425, 3137, \dodoi{10.1111/j.1365-2966.2012.21603.x}

\bibitem[{{Ubach} {et~al.}(2017){Ubach}, {Maddison}, {Wright}, {Wilner}, {Lommen}, \& {Koribalski}}]{ubach17}
---. 2017, \mnras, 466, 4083, \dodoi{10.1093/mnras/stx012}

\bibitem[{{Ueda} {et~al.}(2025){Ueda}, {Andrews}, {Carrasco-Gonz{\'a}lez}, {Guerra-Alvarado}, {Okuzumi}, {Tazaki}, \& {Kataoka}}]{ueda25}
{Ueda}, T., {Andrews}, S.~M., {Carrasco-Gonz{\'a}lez}, C., {et~al.} 2025, arXiv e-prints, arXiv:2507.14443.
\newblock \doarXiv{2507.14443}

\bibitem[{{van der Velden}(2020)}]{cmasher}
{van der Velden}, E. 2020, The Journal of Open Source Software, 5, 2004, \dodoi{10.21105/joss.02004}

\bibitem[{Virtanen {et~al.}(2020)Virtanen, Gommers, Oliphant, Haberland, Reddy, Cournapeau, Burovski, Peterson, Weckesser, Bright, {van der Walt}, Brett, Wilson, Millman, Mayorov, Nelson, Jones, Kern, Larson, Carey, Polat, Feng, Moore, {VanderPlas}, Laxalde, Perktold, Cimrman, Henriksen, Quintero, Harris, Archibald, Ribeiro, Pedregosa, {van Mulbregt}, \& {SciPy 1.0 Contributors}}]{scipy}
Virtanen, P., Gommers, R., Oliphant, T.~E., {et~al.} 2020, Nature Methods, 17, 261, \dodoi{10.1038/s41592-019-0686-2}

\bibitem[{Viscardi {et~al.}(2025)Viscardi, Mac{\'\i}as, Zagaria, Sierra, Jiang, Yoshida, \& Curone}]{viscardi2025dust}
Viscardi, E.~M., Mac{\'\i}as, E., Zagaria, F., {et~al.} 2025, Astronomy \& Astrophysics, 695, A147

\bibitem[{Waskom(2021)}]{seaborn}
Waskom, M.~L. 2021, Journal of Open Source Software, 6, 3021, \dodoi{10.21105/joss.03021}

\bibitem[{{Weintraub} {et~al.}(1989){Weintraub}, {Sandell}, \& {Duncan}}]{weintraub89}
{Weintraub}, D., {Sandell}, G., \& {Duncan}, W.~D. 1989, \apjl, 340, L69, \dodoi{10.1086/185441}

\bibitem[{Wilner(2004)}]{wilner2004imaging}
Wilner, D.~J. 2004, New Astronomy Reviews, 48, 1363

\bibitem[{{Wilner} {et~al.}(2005){Wilner}, {D'Alessio}, {Calvet}, {Claussen}, \& {Hartmann}}]{wilner05}
{Wilner}, D.~J., {D'Alessio}, P., {Calvet}, N., {Claussen}, M.~J., \& {Hartmann}, L. 2005, \apjl, 626, L109, \dodoi{10.1086/431757}

\bibitem[{{Winter} {et~al.}(2024){Winter}, {Benisty}, \& {Andrews}}]{winter24}
{Winter}, A.~J., {Benisty}, M., \& {Andrews}, S.~M. 2024, \apjl, 972, L9, \dodoi{10.3847/2041-8213/ad6d5d}

\bibitem[{{Woitke} {et~al.}(2016){Woitke}, {Min}, {Pinte}, {Thi}, {Kamp}, {Rab}, {Anthonioz}, {Antonellini}, {Baldovin-Saavedra}, {Carmona}, {Dominik}, {Dionatos}, {Greaves}, {G{\"u}del}, {Ilee}, {Liebhart}, {M{\'e}nard}, {Rigon}, {Waters}, {Aresu}, {Meijerink}, \& {Spaans}}]{woitke16}
{Woitke}, P., {Min}, M., {Pinte}, C., {et~al.} 2016, \aap, 586, A103, \dodoi{10.1051/0004-6361/201526538}

\bibitem[{{Xin} {et~al.}(2023){Xin}, {Espaillat}, {Rilinger}, {Ribas}, \& {Mac{\'\i}as}}]{xin23}
{Xin}, Z., {Espaillat}, C.~C., {Rilinger}, A.~M., {Ribas}, {\'A}., \& {Mac{\'\i}as}, E. 2023, \apj, 942, 4, \dodoi{10.3847/1538-4357/aca52b}

\bibitem[{{Zagaria} {et~al.}(2025){Zagaria}, {Facchini}, {Curone}, {Williams}, {Clarke}, {Ribas}, {Tazzari}, {Mac{\'\i}as}, {Booth}, {Rosotti}, \& {Testi}}]{zagaria25}
{Zagaria}, F., {Facchini}, S., {Curone}, P., {et~al.} 2025, arXiv e-prints, arXiv:2507.08797, \dodoi{10.48550/arXiv.2507.08797}

\bibitem[{{Zapata} {et~al.}(2017){Zapata}, {Rodr{\'\i}guez}, \& {Palau}}]{zapata17}
{Zapata}, L., {Rodr{\'\i}guez}, L.~F., \& {Palau}, A. 2017, \apj, 834, 138, \dodoi{10.3847/1538-4357/834/2/138}

\bibitem[{{Zhu} {et~al.}(2019){Zhu}, {Zhang}, {Jiang}, {Kataoka}, {Birnstiel}, {Dullemond}, {Andrews}, {Huang}, {P{\'e}rez}, {Carpenter}, {Bai}, {Wilner}, \& {Ricci}}]{zhu19}
{Zhu}, Z., {Zhang}, S., {Jiang}, Y.-F., {et~al.} 2019, \apjl, 877, L18, \dodoi{10.3847/2041-8213/ab1f8c}

\bibitem[{{Zubko} {et~al.}(1996){Zubko}, {Mennella}, {Colangeli}, \& {Bussoletti}}]{zubko96}
{Zubko}, V.~G., {Mennella}, V., {Colangeli}, L., \& {Bussoletti}, E. 1996, \mnras, 282, 1321

\end{thebibliography}
\bibliographystyle{aasjournal}

\clearpage
\clearpage

\appendix

\section{Model Prescription Choices}

\begin{figure*}[t!]
    \centering
    \includegraphics[width=\linewidth]{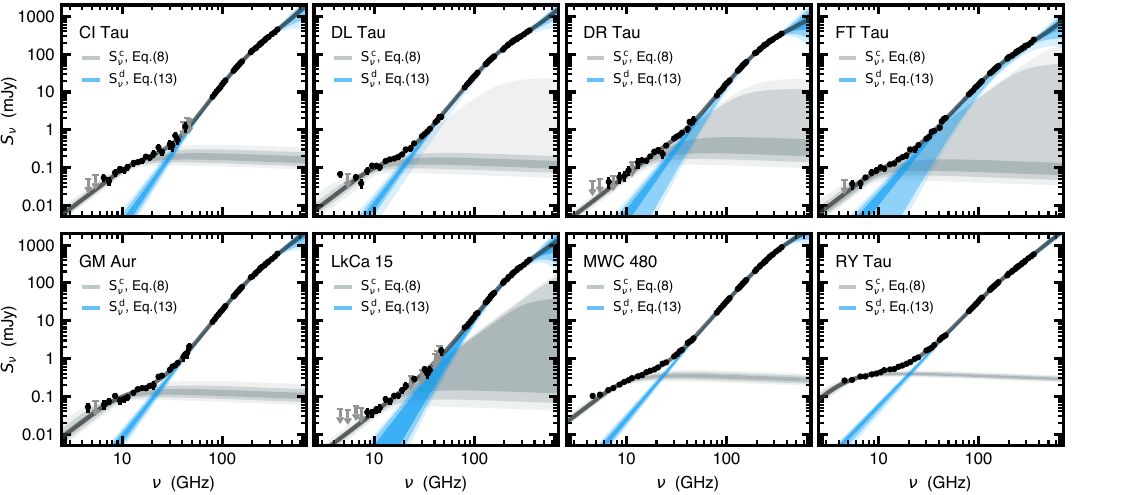}
    \includegraphics[width=\linewidth]{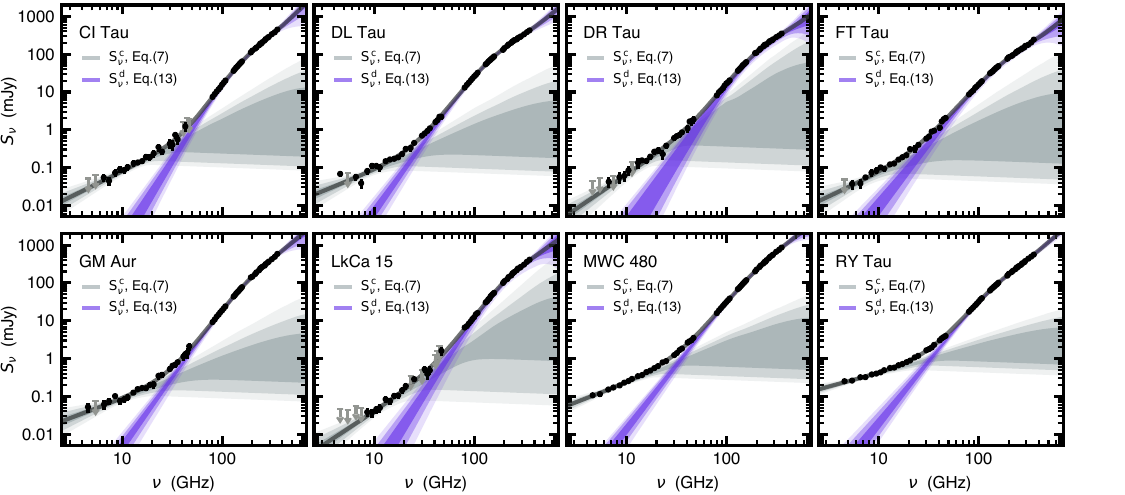}    
    \caption{({\it top rows}) The decomposition of the dust and contamination spectra model posteriors for the fiducial prescriptions used in the main text, with annotations as in Fig.~\ref{fig:fits_gallery}.  The composite spectra posterior medians are marked as gray curves.  ({\it bottom rows}) As above, except for inferences that used the Eq.~(\ref{eq:contam1}) contamination spectrum prescription instead (and the same dust spectrum prescription).}
    \label{fig:ff_disentangle}
\end{figure*}

\begin{figure*}[t!]
    \centering
    \includegraphics[width=\linewidth]{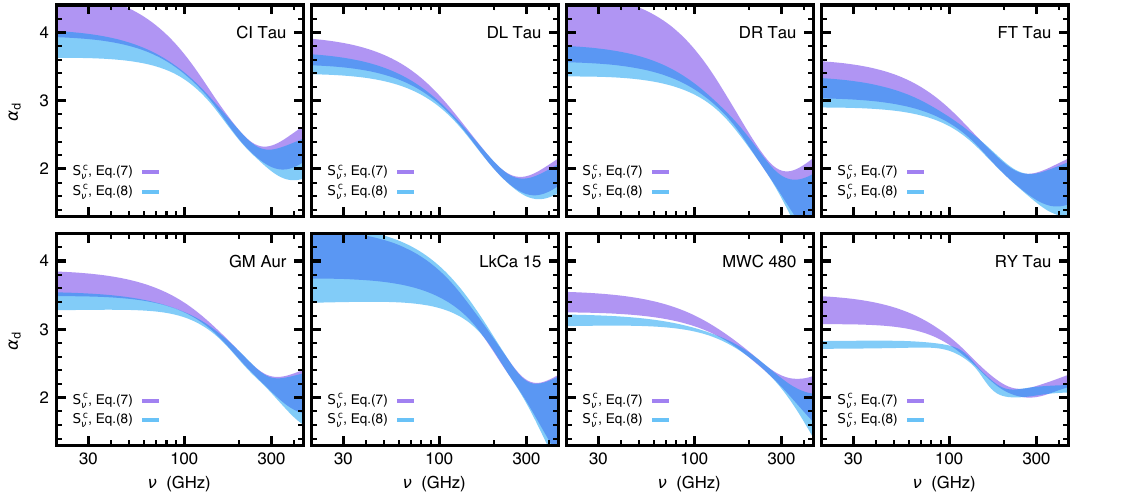}    
    \caption{Direct comparisons of the inferred dust spectral curvatures (the $\alpha_{\rm d}(\nu)$ profiles) for the same dust model prescription (Eq.~\ref{eq:free_index}) and two different assumptions about the contamination prescription (Eq.~\ref{eq:contam1} and \ref{eq:contam2} in purple and blue, respectively).  The shaded regions mark the 68\%\ ($\sim$1-$\sigma$) confidence intervals.  The results are all consistent at the 95\%\ confidence level, but the results for the broken power-law description of the contamination tend toward slightly higher $\alpha_{\rm d}$ ($\sim$0.2--0.3) at low frequencies.}
    \label{fig:ff_alphad}
\end{figure*}

We present here a more extended summary of the spectrum modeling efforts, to help quantify the systematic uncertainties associated with decisions about the modeling prescriptions.  In Section \ref{sec:contam_models}, we considered two options for quantifying the contamination spectra: a simple broken power-law prescription (Eq.~\ref{eq:contam1}) and a more realistic, physically-motivated free-free emission prescription (Eq.~\ref{eq:contam2}).  Though we selected the latter option above, there are small differences associated with the corresponding dust spectra that result from that decision.  Figure \ref{fig:ff_disentangle} compares the inferred spectra for these two options for the same dust spectrum prescription (using Eq.~\ref{eq:free_index}, as above).  The physically-motivated model is in a sense more rigid, which results in stronger constraints (more precision) in most cases.  It is also generally less steep above $\sim$20 GHz, as the emission becomes optically thin.  To compensate for the latter effect, the dust spectra models for the broken power-law prescription tend to be slightly steeper at the lower frequencies, as demonstrated in Figure \ref{fig:ff_alphad}.  That difference is statistically marginal ($\sim$2-$\sigma$ at most, with RY Tau being the most obvious case), with $\alpha_{\rm d}$ differences of $\lesssim$0.2--0.3.  Propagating these results through to the dust mass analysis, we find negligible differences between the constraints from both prescriptions.

\begin{figure*}[t!]
    \centering
    \includegraphics[width=\linewidth]{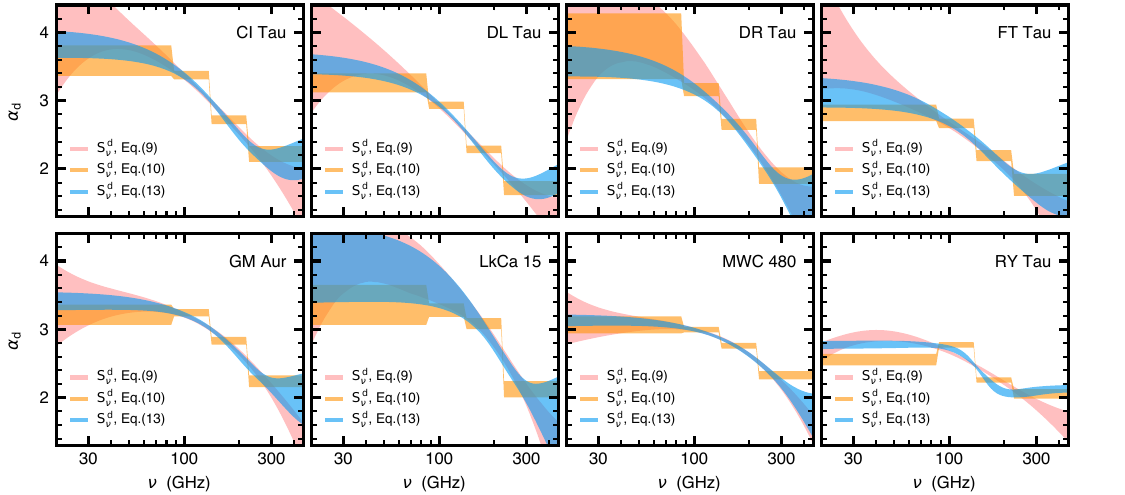}    
    \caption{Analogous comparisons (as in Fig.~\ref{fig:ff_alphad}) of the inferred dust spectral curvatures (the $\alpha_{\rm d}(\nu)$ profiles) for the same contamination prescription (Eq.~\ref{eq:contam2}) and three different assumptions about the dust prescription (Eq.~\ref{eq:poly_exp}, \ref{eq:segs}, and \ref{eq:free_index} in pink, orange, and blue, respectively).  The shaded regions mark the 68\%\ ($\sim$1-$\sigma$) confidence intervals.}
    \label{fig:d_alphad}
\end{figure*}

Analogously, we can compare the inferred model spectra and curvatures for the three different dust spectrum prescriptions (Eq.~\ref{eq:poly_exp}, \ref{eq:segs}, and \ref{eq:free_index}) and a fixed contamination prescription (here we selected Eq.~\ref{eq:contam2}).  The spectra themselves are difficult to distinguish (aside from the polynomial expansion case, because it artificially increases again at frequencies below where the observations probe), and so are not a particularly useful visual diagnostic.  But Figure \ref{fig:d_alphad} compares the inferred curvatures and generally finds strong agreement despite the fundamentally different prescriptions.  The differences are smaller than those found for the different contamination prescriptions, suggesting that disentangling the spectral components is the larger systematic issue in such analyses.

\section{Details on Inferences of Dust Properties}

\begin{figure*}[t!]
    \begin{minipage}[c]{0.5\linewidth}
        \centering
        \includegraphics{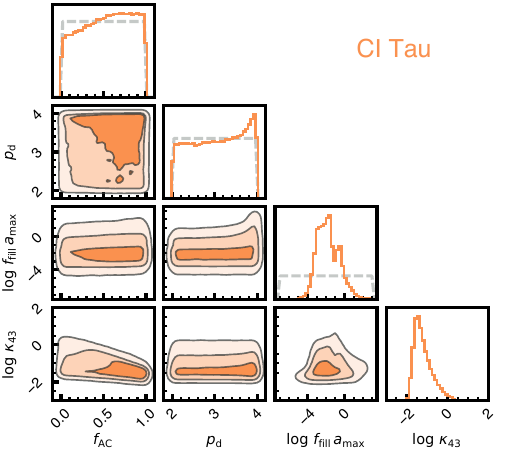}
        \includegraphics{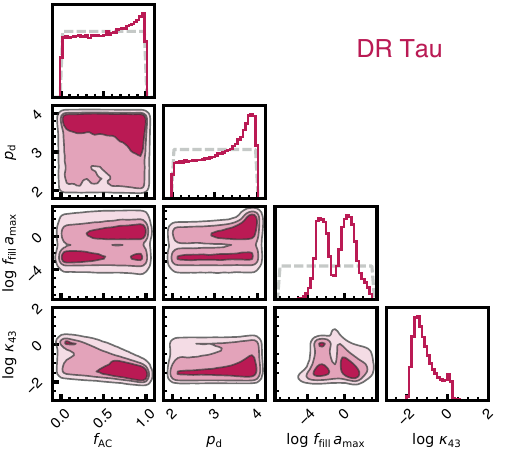}
    \end{minipage}\hfill
    \begin{minipage}[c]{0.5\linewidth}
        \centering
        \includegraphics{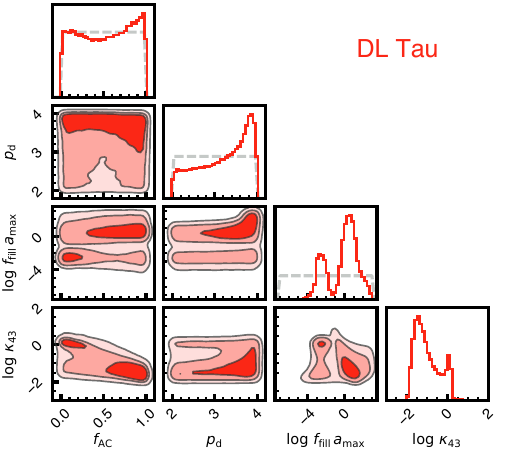}
        \includegraphics{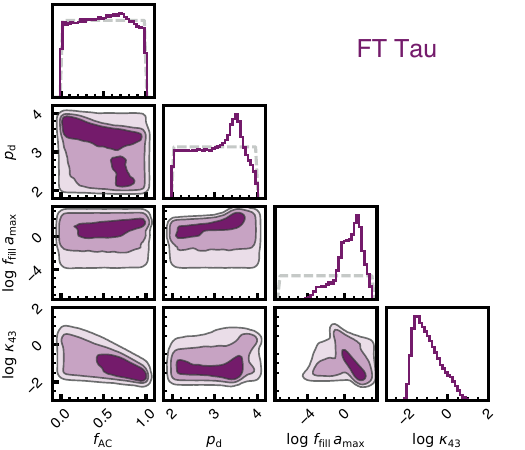}
    \end{minipage}
    \caption{The inferred pairwise marginalized posterior distributions for the dust properties and opacities determined in Sect.~\ref{sec:Mdust}.  Contours are drawn at 1, 2, and 3-$\sigma$ confidence intervals from dark to light shadings.  The adopted prior distributions are marked as dashed gray curves in the one-dimensional marginalized distributions along the diagonals.}
    \label{fig:corner1}
\end{figure*}

\begin{figure*}[t!]
    \begin{minipage}[c]{0.5\linewidth}
        \centering
        \includegraphics{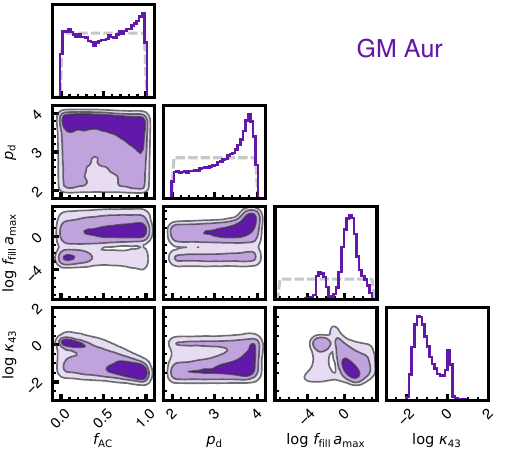}
        \includegraphics{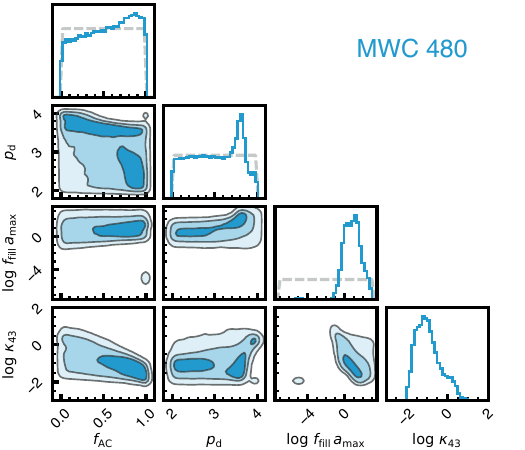}
    \end{minipage}\hfill
    \begin{minipage}[c]{0.5\linewidth}
        \centering
        \includegraphics{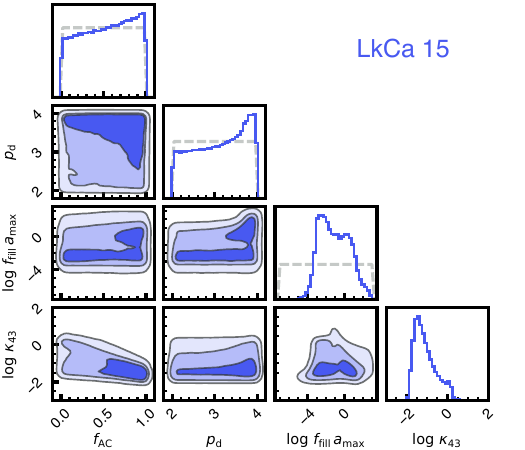}
        \includegraphics{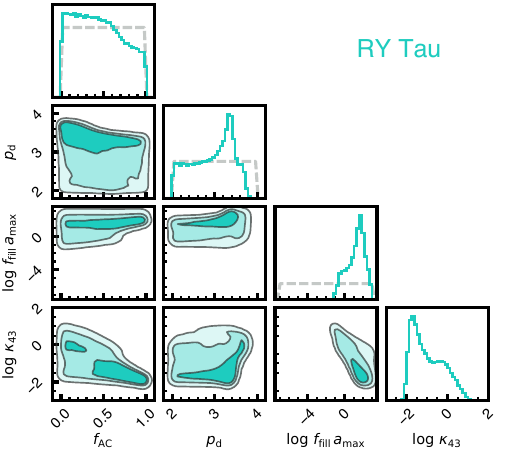}
    \end{minipage}
    \caption{As in Fig.~\ref{fig:corner1}, for the remaining targets.}
    \label{fig:corner2}
\end{figure*}

In Section \ref{sec:Mdust}, we inferred constraints on a set of dust properties for each disk conditioned on the measured 43 GHz dust spectral indices.  For the sake of completeness, we summarize those constraints with marginalized pairwise covariance plots in each case in Figures \ref{fig:corner1} and \ref{fig:corner2}.  Because of the correlation between the filling factor $f_{\rm fill}$ and maximum particle size $a_{\rm max}$, we consolidate to show the product of those parameters \citep[e.g., see][]{kataoka14}.  While there are no clear constraints on the dust size distribution indices or compositions (amorphous carbon fractions), but these covariance plots relate information on how the $f_{\rm fill} \times a_{\rm max}$ constraints relate to the 43 GHz absorption opacities.  

\end{document}